\documentclass[]{aa} 
\usepackage{graphicx}
\usepackage{longtable}
\usepackage{txfonts}

\usepackage{natbib} 

\begin{document}

\title{Star-planet magnetic interaction 
and activity  in late-type stars with close-in planets }

   \subtitle{}

   \author{A.~ F.~ Lanza
          }

\offprints{A. ~F. ~Lanza}  
\titlerunning{Close-in planets and magnetic activity}
    \authorrunning{A.~ F.~ Lanza}

   \institute{INAF-Osservatorio Astrofisico di Catania, Via S. Sofia, 78 -- 95123 Catania, Italy\\
              \email{nuccio.lanza@oact.inaf.it}
             }

   \date{Received ...; accepted ...}

 
  \abstract
   {Late-type stars interact with their close-in planets through their coronal magnetic fields. }
   {We introduce a theory for the interaction between the stellar and planetary fields focussing on the processes that release magnetic energy in the stellar coronae. }
   {We consider the energy dissipated by the reconnection between the stellar and planetary magnetic fields as well as that made available by the modulation of the magnetic helicity of the coronal field produced by the orbital motion of the planet. We   estimate the  powers released by both processes in the case of axisymmetric and non-axisymmetric, linear and non-linear force-free coronal fields finding that they scale as $B_{0}^{4/3} B_{\rm p0}^{2/3} R_{\rm p}^{2} v_{\rm r}$, where $B_{0}$ is the mean stellar surface field, $B_{\rm p0}$ the planetary field at the poles, $R_{\rm p}$ the radius of the planet, and $v_{\rm r}$ the relative velocity between the stellar and the planetary fields.   }
   {A chromospheric hot spot or a flaring activity phased to the orbital motion of the planet are found only when the stellar field is axisymmetric. In the case of a non-axisymmetric field, the time modulation of the energy release is multiperiodic and can be easily confused with the intrinsic stellar variability.
   We apply our theory to the systems with some reported evidence of star-planet magnetic interaction finding a dissipated power at least one order of magnitude smaller than that emitted by the chromospheric hot spots. The phase lags between the planets and the hot spots are reproduced by our models in all the cases except for $\upsilon$~And. }
   { The chromospheric hot spots rotating in phase with the  planets cannot be explained by the energy dissipation produced by the interaction between stellar and  planetary fields as considered by our models and require a different mechanism.  }

   \keywords{stars: planetary systems -- stars: activity -- stars: late-type -- stars: magnetic fields -- stars: individual (HD~179949, HD~189733, $\upsilon$~Andromedae, $\tau$~Bootis)
               }

   \maketitle
%

\section{Introduction}

Close-in planets interact with their host stars through tides and magnetic fields. Here we focus on the interaction between the coronal field of a late-type star and a planet orbiting within its outer corona. \citet{Shkolniketal05,Shkolniketal08}  presented evidence of a chromospheric hot spot that moves in phase with the orbit of the planet rather than with the stellar rotation period in the cases of HD~179949 and $\upsilon $~Andromedae. The hot spot was not present in all the observing seasons and its maximum visibility was shifted with respect to the planet inferior conjunction. The non-steadiness of the phenomenon was confirmed also by \cite{Poppenhaegeretal11} who did not detect any evidence for star-planet magnetic interaction (hereafter SPMI) when re-observing $\upsilon$~And in 2009. The night-to-night variation of the chromospheric emission of some stars with hints of SPMI appears to be correlated with the ratio $M_{\rm p} \sin i/P_{\rm orb}$ of the projected mass of the planet to the orbital period. This ratio can be assumed as a measure of the planetary magnetic moment by extrapolating the correlation observed in the Solar System between the magnetic moment and the spin angular momentum of  the magnetized bodies \citep{Shkolniketal08}. Therefore, the proposed correlation suggests that stellar activity is somehow affected by the field of a close-by planet. 
{ However, \citet{Milleretal12} did not find evidence of activity phased with the planet  in the case of WASP-18 whose massive ($M=10.4$ Jupiter masses) and close-in (orbital semimajor axis 0.02~AU) companion is expected to induce a much larger effect than in the cases of HD~179949 or $\upsilon$~And. A possible explanation could be the weak magnetic field of the star, as indicated by its remarkably low level of activity, that makes the interaction undetectable in spite of the supposedly strong planetary field.} 

Considering a sample of  stars with planets, \citet{Cantomartinsetal11} did not find any difference in the mean level of their chromospheric emission in comparison with a sample of late-type stars without detected planets, while \citet{Gonzalez11} claims that planet-hosting stars are, on the average, slightly less active than stars without planets. An intriguing result was recently obtained by \citet{Hartman10} who found a correlation between the chromospheric emission and the surface gravity of the planet in a sample of stars hosting hot Jupiters. 

The search for an SPMI signature in the X-ray coronal emission  led to a controversy. \citet{Kashyapetal08} found that stars with close-in 
(orbital semi-major axis $a < 0.15$~AU) massive planets have an X-ray flux $\approx 2$ times higher than stars with distant planets ($a > 1.5$~AU) and \citet{Scharf10} found a correlation between X-ray luminosity and the mass of the exoplanets, suggesting  that it could be used to infer the relative  intensity of the planetary magnetic field if the excess energy was released by the reconnection between the stellar field and the planetary field. However, \citet{Poppenhaegeretal10} and \citet{PoppenhaegerSchmitt11}, investigating a complete sample of  planet-hosting stars within 30~pc, did not find any correlation between stellar activity and orbital semimajor axis or planet mass that cannot be traced back to selection effects. 

Another possibility concerns a flaring coronal activity modulated by the orbital motion of the planet, as recently suggested in the case of HD~189733 by \citet{Pillitterietal11}. Finally, the possibility that SPMI manifests itself in the starspot activity observed in the photosphere has  been also considered and is discussed in detail by \citet{Lanza11}.

Several models have been proposed to account for the possible signatures of SPMI.  \citet{Preusseetal06} and \citet{Koppetal11} considered the impact on the chromosphere of  Alfven waves excited by the orbital motion of the planet. Such waves can reach the surface of the star because the planet is located within the region where the Alfven velocity is larger than the speed of the stellar wind so a magnetic perturbation can move back to the star instead of being blown away by the  wind as happens in the case of Jupiter or Saturn in the Solar System since they are located in the region of the heliosphere where the solar wind is super-alfvenic. These models can account for the phase lag between the planet and a chromospheric hot spot by assuming that the spot is due to  the dissipation of the waves  where they impact onto the chromosphere,  but it is very difficult to account for the power radiated by the hot spots  that is of the order of $\sim 10^{20}$~W \citep{Shkolniketal05} because the wave energy is not focussed onto the star. Other models have addressed the interpretation of the phase lag  \citep{McIvoretal06,Lanza08} or  the mechanism of chromospheric heating \citep[e.g., ][]{GuSuzuki09}. 

The problem of the energy budget of the interaction has been considered in numerical simulations, starting from the pioneering work by \citet{Ipetal04} who, however, adopted coronal magnetic fields and relative orbital velocity of the planet that appear to be too large by about one order of magnitude than in the cases of HD~179949.
The sophisticated MHD simulations of \citet{Cohenetal09,Cohenetal11a,Cohenetal11b} are much more realistic. 
{Specifically, \citet{Cohenetal11b} simulated the magnetic environment around HD~189733, a K dwarf hosting a hot Jupiter, approximating the large-scale variation of the radial magnetic field of the star at the photosphere as derived by \citet{Faresetal10}. They found that the power released by the star-planet magnetic interaction  is sufficient to account for the flaring activity detected by \citet{Pillitterietal11} or the chromospheric hot spots observed by \citet{Shkolniketal05} in similar systems. 

More recently, \citet{Vidottoetal12} have simulated the time variations of the magnetized wind of $\tau$~Bootis by extrapolating the radial photospheric field of the star as mapped by \citet{Catalaetal07}, \citet{Donatietal08}, and \citet{Faresetal09} at different phases of the stellar activity cycle. The interaction of a strongly magnetized pre-main-sequence star with a close-in massive planet has been simulated  by \citet{Vidottoetal09,Vidottoetal10}. They also derived  the power released by magnetic reconnection between the stellar and  planetary fields, that could reach $\approx 5 \times 10^{19}$~W for a planet orbiting at a distance of $7-10$ stellar radii, and  estimated the expected radio emission flux and  the effect of the strong stellar field on the migration of  the planet.

Other relevant work includes the potential magnetic field model developed by \citet{Adamsetal11} to describe the magnetic interaction of the components of close pre-main-sequence binary stars. A variant of the model has been applied  to describe the interaction between the coronal field of a host star and the magnetized outflow of its close-in evaporating planet \citep{Adams11}. }

{In spite of these works,} a general theory of the SPMI capable of accounting for the required irradiated power is still lacking. A first step in this direction has been the work by \citet{Lanza09} who found that the power  dissipated by magnetic reconnection at the boundary surface between the planetary magnetosphere and the stellar coronal field is of the order of $10^{17}$~W, i.e., insufficient by at least three orders of magnitude, but the interaction with the planetary field may trigger the dissipation of the magnetic energy over a much larger volume by decreasing the helicity of the coronal field (see Sect.~\ref{model} for details). In the present paper, we explore this mechanism in more detail {considering the magnetic flux tube of the stellar coronal field that interacts with the magnetosphere of the planet} and provide  analytic formulae to estimate the dissipated power together with some illustrative applications to the systems that have shown some evidence of SPMI. Note that most of our considerations can be extended to the case of close binary systems with a stellar \citep[see, e.g., ][]{Adamsetal11,Getmanetal11,Strassmeieretal11} or a brown dwarf secondary component \citep[e.g., ][]{Lenzetal10}, although here we shall address only the case of close-in planets. 

\section{The model}
\label{model}

\subsection{A magnetostatic coronal model}
\label{coronal_model}

{
We need a model of the corona of the host star to test the validity of our assumptions on the energetics of the star-planet interaction. Assuming that the coronal plasma is in hydrostatic equilibrium under the action of the pressure gradient, the gravity, and the Lorentz force, we have:
\begin{equation}
-\nabla p + \rho \nabla \Phi + {\vec J} \times {\vec B}= 0,
\label{hydro_eq}
\end{equation}
where $p$ is the plasma pressure, $ \rho$ its density, $\Phi$ the total potential (gravitational plus centrifugal), ${\vec B} = B \hat{\vec s}$ the magnetic field, with $\hat{\vec s}$ the unit vector in the direction of the magnetic field, and $\vec J$ the current density. The centrifugal potential can be neglected in our case, provided that the star has a rotation period of at least $\sim 10$ days  and a radius and mass comparable with those of the Sun. Therefore, $\Phi \simeq  G M /r$, where $G$ is the gravitation constant, $M$ the mass of the star, and $r$ the distance from its centre. The thermal conduction is high along magnetic field lines and strongly inhibited in the orthogonal direction \citep[cf., e.g., ][]{Priest84}, thus we assume that the temperature $T$ is constant along magnetic field lines, i.e., $\partial T / \partial s = 0$. The dot product of Eq.~(\ref{hydro_eq}) by the unit vector $\hat{\vec s}$ gives:  
\begin{equation}
-\frac{\partial p}{\partial s} + \rho \frac{\partial \Phi}{\partial r} (\hat{\vec r} \cdot \hat{\vec s}) = 0.
\end{equation}
Since $ dr = (\hat{\vec r} \cdot \hat{\vec s}) ds$, this becomes: 
\begin{equation}
-\frac{\partial p}{\partial r} + \rho \frac{\partial \Phi}{\partial r} = 0, 
\end{equation}
that can be integrated along a given magnetic field line by considering the   ideal gas law to eliminate $\rho = \tilde{\mu}p / \tilde{R} T$, where $\tilde{\mu}$ is the mean molecular weight of the coronal plasma, $\tilde{R}$ the gas constant, and $T$ the temperature along the given field line. Thus, the variation of the pressure along a field line is:
\begin{equation}
p(r) = p_{0} \exp \left[- \left( \frac{R}{H_{0}} \right) \left( 1 - \frac{R}{r} \right)\right], 
\label{pressure_run}
\end{equation}
 where $R$ is the radius at base of the corona, that we assume to coincide with the radius of the star, and $H_{0} = (\tilde{R} T R ^{2})/ (G \tilde{\mu }  M)$ the pressure scale height at the base of the corona. For the Sun, $\hat{\mu}=0.6$ and $H_{0} = 5.1 \times 10^{7} T$~m, with $T$ in MK.  The variation of the density can be derived from the ideal gas law and has the same radial dependence of the pressure along a given field line. Note, however,  that $T$, $p_{0}$, and the base density $\rho_{0}$  vary in general from one field line to the other. We shall apply this simple model in Sect.~\ref{coronal_param} to justify the assumptions of our energy dissipation model. 

}
\subsection{Power dissipated by magnetic reconnection}
\label{reconnection}

The orbital motion of the planet inside the stellar corona produces a continuous reconnection between the coronal magnetic field lines and the planetary field lines. \citet{Lanza09} studied this phenomenon and estimated the power dissipated in the stellar corona. We briefly recall the main assumptions and the results of that investigation that are useful for the present study. We assume that the planet is on a circular orbit located on the equatorial plane of the star. The surface of the planetary magnetosphere, where its field lines interact with those of the coronal field, is assumed to be a sphere of radius $R_{\rm m}$. As a matter of fact, the magnetospheric boundary can be elongated in the direction of the orbital motion of the planet \citep[cf., e.g., ][]{Cohenetal11b}. However,  we specialize our theory to the case when the orbital velocity  of the planet is much smaller than the Alfven velocity in the stellar corona. Therefore, the magnetic field configuration can be regarded as  magnetostatic and the boundary of the magnetosphere is nearly spherical because it is defined by the balance between the magnetic pressure of the coronal field and the pressure of the planetary field  assumed to be a dipole. Assuming that the magnetic pressure of the coronal field $ \vec B ({\vec r}_{\rm m}) $ at the boundary of the magnetosphere is in equilibrium with the pressure of the planetary field ${\vec B}_{\rm p} ({\vec r}_{\rm m})$, we have:
\begin{equation}
{B}^{2}_{\rm p} ({\vec r}_{\rm m}) = {B}^{2} ({\vec r}_{\rm m}), 
\label{magnetosphere_eq}
\end{equation}
where ${ \vec r}_{\rm m}$ is the position vector of a generic point on the boundary of the magnetosphere. The planetary field can be assumed to be that of a dipole, so that its variation with the distance $\Delta$ from the centre of the planet is:
\begin{equation}
B_{\rm p} = B_{\rm p0} \left( \frac{\Delta}{R_{\rm p}} \right)^{-3},
\label{planet_dip}
\end{equation}
where $B_{\rm p0}$ is the field at the poles of the planet and $R_{\rm p}$ its radius. At the boundary of the magnetosphere, we have $\Delta = R_{\rm m}$ and, considering Eq.~(\ref{magnetosphere_eq}), we find:
\begin{equation}
R_{\rm m}  = R_{\rm p} \left[ \frac{B({\vec r}_{\rm p})}{B_{p0}} \right]^{-1/3}, 
\label{rm}
\end{equation}
where $B ({\vec r}_{\rm p}) $ is the intensity of the ambient coronal field at the position ${\vec r}_{\rm p}$ of the planet. Since the radius of the magnetosphere is generally small in comparison with the lengthscale of variation of $B$ in the outer stellar corona, we neglect the variation of $B$ across the magnetosphere and consider its intensity at the location of the planet ${\vec r}_{\rm p}$. 

The power released by the magnetic reconnection at the boundary of the planetary magnetosphere can be estimated as:
\begin{eqnarray}
P_{\rm rec} & \simeq & \gamma_{\rm rec} \frac{\pi}{\mu} B^{2}({\vec r}_{\rm p}) R_{\rm m}^{2} v_{\rm rel} = \nonumber \\
 & = & \gamma_{\rm rec} \frac{\pi}{\mu} R_{\rm p}^{2} B^{4/3}({\vec r}_{\rm p}) B_{\rm p0}^{2/3} v_{\rm rel}, 
\label{rec-power}
\end{eqnarray}
where $\mu $ is the magnetic permeability of the plasma, $0 < \gamma_{\rm rec} < 1$ an efficiency factor that depends on the angle between the interacting magnetic field lines \citep[e. g., ][]{Priest03}, $v_{\rm rel}$  the relative velocity between the reconnecting magnetic field lines, i.e., the planetary and the stellar fields, and we have assumed that an effective surface $\pi R_{\rm m}^{2}$ is available for the interaction of the reconnecting field lines. 

The power dissipated in the reconnection process has been estimated by \citet{Lanza09} to be of the order of $10^{17}$~W in the case of a stellar dipolar potential field, i.e., insufficient by $\sim 3$ orders of magnitude. In Sect.~\ref{results}, we shall see that considering  force-free magnetic fields we can increase that power by about two orders of magnitude in the most extreme cases. This is still insufficient to account for the power irradiated by the chromospheric hot spots. Therefore, we shall consider another mechanism that can release greater powers by extracting energy from a larger coronal volume, not only from that taking part in the reconnection. This mechanism is connected with the role of magnetic helicity in stellar coronae that we shall briefly describe in the next Section.

\subsection{The role of magnetic helicity in stellar coronae}
\label{helicity}

We assume that the Lorentz force is dominating over all the other forces in the region of the stellar corona where a close-in planet is  located. { This is valid provided that the ratio between the plasma pressure and the magnetic pressure $\beta \equiv 2\mu p/B^{2}$ is much smaller than the unity in the considered coronal domain, as we shall show in Sect.~\ref{coronal_param}. } With this assumption, the force-free approximation can be applied to describe the coronal field, i. e.,
\begin{equation}
\nabla \times {\vec B} = \alpha {\vec B},
\label{force-free-def}
\end{equation}
 where the force-free parameter $\alpha$ is constant along each field line, as immediately follows from the curl of the defining equation (\ref{force-free-def}). In general, $\alpha$ will vary from one field line to the other (non-linear force-free field). When it is constant for all the field lines, the field is said to be a linear force-free field. We assume that the orbital motion of the planet and the associated plasma flow do not significantly perturb the stellar field configuration which is determined by the boundary conditions set at the base of the corona on the stellar surface. If those boundary conditions evolve on a timescale much longer than the Alfven travel time across the corona, the field can be assumed to be at each instant  in a magnetostatic configuration as described by Eq.~(\ref{force-free-def}). {Of course, the force-free approximation is not valid at the stellar photosphere, where the pressure of the plasma is comparable or  greater than the magnetic pressure, but we can assume that our low-beta approximation is valid starting from the base of the corona that we assume for simplicity to coincide with the stellar surface because the photospheric and chromospheric pressure scale heights are much smaller than the radius of the star. Note also that the flux systems of the stellar corona and of the planetary magnetosphere are topologically separate because the planetary magnetic field is potential close to the surface of the planet. Therefore, a stationary magnetic field line that interconnects the planetary field with the stellar coronal field would have a zero value of $\alpha$ that is in general incompatible with the presence of electric currents flowing through the stellar corona. In other words, the field lines of the planetary field must be confined within the planetary magnetosphere and interact with the stellar field lines only on  the boundary of the  magnetosphere where a time-dependent reconnection occurs. A large-scale flux system interconnecting the stellar and planetary fields in a low-beta regime is possible only when the stellar coronal field is assumed to be potential as in the models of \citet{Adams11} and \citet{Adamsetal11}. }
 
 If the magnetic field is confined to some closed volume $V$, i.e., its field lines do not cross the boundary $S$ of $V$, it is possible to define a conserved quantity in ideal MHD called magnetic helicity $H$:
 \begin{equation}
 H =\int_{V} {\vec A} \cdot {\vec B} \ dV, 
 \label{h-def}
 \end{equation}
 where $\vec B = \nabla \times {\vec A}$ is the magnetic field and $\vec A$ its vector potential. $H$ is a topological measure of the twisting of the magnetic field lines and of their degree of cross linkage \citep[cf. ][ for more details]{BergerField84,Demoulinetal06}. The conservation of the helicity means that, if the magnetic Reynolds number is very large, so that the diffusion of the field is negligible in comparison with the induction effects, the value of $H$ does not change vs. time whatever the plasma motions  inside the volume $V$. Moreover, the minimum energy state allowed in the case of a finite helicity is a linear force-free field {\citep{Woltjer58}}, instead of a potential  field which would represent the absolute minimum that can be reached only when the helicity is zero \citep[cf., e.g.,][ Ch.~3]{Priest84}. 
 
In a non-ideal plasma, the finite resistivity produces a decay of the field with a conversion of the magnetic energy into thermal and kinetic energies. Experiments with laboratory plasma and theoretical considerations have shown that the variation of the magnetic helicity during the relaxation to the minimum energy state is extremely slow in comparison with the  magnetic energy dissipation \citep[e.g., ][]{Berger84}; in practice, the relaxation proceeds under the constraint determined by the helicity conservation and the final relaxed state is a linear force-free field with the same amount of helicity as the initial field {\citep{Taylor74,Taylor86}}.
  
The general results recalled above can be applied to a stellar corona if we account for the differences between a laboratory plasma and a coronal plasma.  Firstly, the coronal field lines are not confined by a magnetic surface, i.e., a surface over which the normal component of the field vanishes. They cross the surface of the star so that the magnetic helicity is no longer a  gauge invariant quantity, i.e., it varies by adding   the gradient of a generic scalar function $\chi$ to the vector $\vec A$ leading to the gauge transformation $\vec A \rightarrow {\vec A} + \nabla \chi$. To overcome this difficulty,  \citet{BergerField84} and \citet{Berger85} defined a relative magnetic helicity 
that is independent of the gauge assumed for the vector potential and is conserved in ideal MHD. In a finite domain, the minimum energy field having a given relative helicity is again a linear force-free field  \citep[cf.  Sect.~III of ][]{Berger85}.  
The second difference between a laboratory plasma and a coronal plasma is that the coronal field can extend to the infinity. In this case the minimum energy state is actually reached by driving the twist of the field lines to the infinity which dilutes the magnetic helicity density until the final  state is virtually indistinguishable from the potential field corresponding to the boundary conditions applied at the coronal base \citep[cf., e.g., ][ and references therein]{Veksteinetal93,Zhangetal06}.
However, a stellar corona consists also of closed magnetic structures that are confined by overlying open fields, as in the case of a magnetic arcade with an helmet streamer on its top. In those closed structures, the accumulation of magnetic energy due to the shearing motions of the photospheric footpoints of the field lines is accompanied by a relaxation produced by the dissipation of magnetic energy localized in thin current sheets. The minimum energy state {of each confined structure} is predicted to be a linear force-free field satisfying the constraint of  relative helicity conservation {\citep[see ][ for  details]{Dixonetal89}}. However, if the amount of helicity accumulated in a confined structure exceeds a certain threshold, that depends on the boundary conditions, the magnetic configuration may become unstable and erupts producing a coronal mass ejection (hereafter CME) that takes away most of the accumulated helicity allowing the remaining field to relax to a quasi-potential state \citep[cf. ][]{Zhangetal06,ZhangFlyer08,Milleretal09}.

{ The processes that dissipate the excess magnetic energy and lead to the linear force-free minimum energy state are outside of the scope of the  theory described above. In general, they involve the generation of hydromagnetic instabilities and/or the formation of localized currents  under the action of photospheric motions or the emergence of new magnetic flux \citep[e.g., ][]{Priestetal05,Browningetal08}. Since the timescale for attaining that minimum energy state is of particular relevance to our model, we briefly refer to the work of \citet{Browningetal08} who suggest that the non-linear developments of ideal magnetohydrodynamic instabilities are the best candidates for the dissipation of the  excess energy in confined structures.  The reason for choosing ideal rather than resistive instabilities is that only the former have sufficiently short timescales in the highly conducting coronal plasma to be relevant to flares or coronal heating. Their 3D numerical simulations of the non-linear evolution of  a kink instability in a magnetic flux tube show that the field is driven toward the minimum energy state while its relative helicity is conserved with a characteristic time scale comparable with the Alfven transit time from one base of the flux tube to the other. This compares well with the observations of solar flares showing that most of the energy released in the impulsive phase is dissipated over timescales of the order of $10-100$~s in flux tubes having lengths ranging from $ 10^{7}$ to $10^{8}$~m, i.e., their relaxation occurs on timescales comparable with the Alfven transit time along the field lines since the typical Alfven velocity in the solar corona is of the order of $10^{6}$~m~s$^{-1}$. 
}

\subsection{Assumptions on the characteristic timescales}
{
We shall consider the variation of the relative helicity produced by the orbital motion of the planet across the coronal field of its host star and estimate the energy made available for dissipation by this process.
 The application of the  above theory  is considerably simplified when the timescale $t_{\rm H}$ for the helicity variation  is considerably longer than the Alfven travel time $t_{\rm A}$ along the magnetic field lines from the star to the planet. On the other hand, the timescale $t_{\rm R}$ for the relaxation to the minimum energy state under the constraint of helicity conservation can be assumed to be comparable with $t_{\rm A}$ (cf. the final paragraph of Sect.~\ref{helicity}). Therefore,  we can assume that the coronal field is always close to the minimum energy state determined by  its instantaneous relative helicity. 
 These hypotheses allow us to  study the evolution of the coronal field as a sequence of 
 magnetostatic configurations to which a simple model for the energy dissipation  can be applied (see Sects.~\ref{helicityandenergy} and~\ref{h-variation}). }  

\subsection{Connection between helicity and magnetic energy}
\label{helicityandenergy}

The relative helicity of a coronal magnetic field configuration can be computed following the method described in \citet{BergerField84}. Specifically, we consider a magnetic flux tube connecting the surface of the star with the magnetosphere of the planet (cf. Fig.~\ref{magnetosphere_sketch}). Its bases are $F_{1}$ on the surface and $F_{2}$ on the magnetospheric boundary. Using the formulation of \citet{Demoulinetal06}, its relative magnetic helicity can be written as:
\begin{equation}
H_{\rm R} = \int_{V_{\rm f}} ({\vec A} + {\vec A}_{\rm p}) \cdot ({\vec B} - {\vec B}_{\rm p}) \ dV,
\label{hrel-def}
\end{equation}
where $V_{\rm f}$ is the volume of the magnetic flux tube, ${\vec B} = \nabla \times {\vec A}$ the magnetic field, ${\vec A}$ its vector potential, ${\vec B}_{\rm p} = \nabla \times {\vec A}_{\rm p}$ the potential magnetic field having the same normal component of ${\vec B}$ on the boundary of $V_{\rm f}$, and ${\vec A}_{\rm p}$ its vector potential (see Appendix~\ref{App1}).  We adopt the so-called Coulomb gauge, i.e., $\nabla \cdot {\vec A} = 0$ and $\nabla \cdot {\vec A}_{\rm p} = 0$. Applying  vector calculus identities and  Gauss' theorem, Eq.~(\ref{hrel-def}) can be recast as:
\begin{equation}
H_{\rm R} = \int_{V_{\rm f}} {\vec A} \cdot {\vec B} \ dV + \oint_{S_{\rm f}} ({\vec A} \times {\vec A}_{\rm p} - \psi {\vec A}_{\rm p} ) \cdot \hat{\vec n} \ dS,
\label{hreldef2}
\end{equation}
where $S_{\rm f}$ is the closed surface bounding the volume $V_{\rm f}$, $\hat{\vec n}$ the unit outward normal to $S_{\rm f}$, and 
$\psi $ the scalar potential of ${\vec B}_{\rm p}$, i.e., ${\vec B}_{\rm p} = \nabla \psi $. 

The connection between the magnetic energy and the relative helicity can be derived by considering that $\nabla \cdot ({\vec A} \times {\vec B}) = {\vec B} \cdot (\nabla \times {\vec A}) - {\vec A} \cdot (\nabla \times {\vec B}) = B^{2} - \alpha ({\vec A} \cdot {\vec B})$, where we have applied the force-free condition $\nabla \times {\vec B} = \alpha {\vec B}$ to the field of the flux tube. Since $\alpha$ varies only across the field lines, the variation of $\alpha$ across the flux tube section is limited, i.e., $\alpha_{\rm min} < \alpha < \alpha_{\rm max}$, with the minimum and maximum values close to each other if the cross section is sufficiently small. The magnetic  energy $E$ of the flux tube  is:
\begin{eqnarray}
E \equiv \int_{V_{\rm f}} \frac{B^{2}}{2\mu} &  = &  \frac{1}{2\mu}  \int_{V_{\rm f}}  \alpha ({\vec A} \cdot {\vec B}) \ dV + \frac{1}{2\mu} \int_{V_{\rm f}}  \nabla \cdot ({\vec A} \times {\vec B}) \ dV  = \nonumber \\ 
  & = & \frac{1}{2\mu} \int_{V_{\rm f}}  \alpha ({\vec A} \cdot {\vec B}) \ dV + \frac{1}{2\mu} \oint_{S_{\rm f}}  ({\vec A} \times {\vec B}) \cdot \hat{\vec n} dS,
\end{eqnarray}
where  the second equality follows by the application of  Gauss' theorem. Applying the mean-value theorem to the first integral, we find:
\begin{eqnarray}
E  &  = & \frac{\langle \alpha \rangle}{2 \mu} \int_{V_{\rm f}} {\vec A} \cdot {\vec B} \ dV + \frac{1}{2\mu} \oint_{S_{\rm f}}  ({\vec A} \times {\vec B}) \cdot \hat{\vec n} dS = \nonumber \\
 & = & \frac{\langle \alpha \rangle}{2 \mu} H_{\rm R} +  \frac{1}{2\mu} \oint_{S_{\rm f}}  [\langle \alpha \rangle (\psi {\vec A}_{\rm p} -{\vec A}\times {\vec A}_{\rm p}) + {\vec A} \times {\vec B}] \cdot \hat{\vec n} dS,
 \label{eqener}
\end{eqnarray}
where the mean value $\alpha_{\rm min} < \langle \alpha \rangle < \alpha_{\rm max}$ and we have made use of 
Eq.~(\ref{hreldef2}) to transform the first integral on the r.h.s. This expression generalizes
Eq.~(17) of \citet{Berger85} to the case of a non-linear force-free field.
Since the magnetic field inside the flux tube occupies a finite volume and is confined by the closed surface $S_{\rm f}$, its minimum energy state is the linear force-free field satisfying the boundary conditions and the constraint set by the conservation of the relative helicity.
If the dissipation of the excess magnetic energy is fast in comparison with the helicity variation, we can assume that the field is always  close to such a minimum energy state. A variation of the relative helicity or of the boundary conditions that changes the surface integral in the r.h.s. of Eq.~(\ref{eqener}) then produces a variation of the energy of the field.  The contribution of the surface integral to the energy can be written in terms of the two quantities:
\begin{eqnarray}
\Theta & \equiv &  \oint_{S_{\rm f}}  (\psi {\vec A}_{\rm p} - {\vec A} \times {\vec A}_{\rm p}) \cdot \hat{\vec n} dS, \mbox{ and }\\
\Sigma & \equiv & \frac{1}{2\mu} \oint_{S_{\rm f}}  ({\vec A} \times {\vec B}) \cdot \hat{\vec n} dS, 
\end{eqnarray}
We focus on the effects produced by the orbital motion of the planet that changes the boundary conditions on the base $F_{2}$. 
The variation of the surface terms is confined to  $F_{2}$ because we  assume that the stellar coronal field is not perturbed outside  the planetary magnetosphere by the motion of the planet.  This gives:
\begin{equation}
\frac{dE}{dt} = \frac{\langle \alpha \rangle}{2 \mu} \left( \frac{dH_{\rm R}}{dt} +  \frac{d \Theta}{dt}  \right) + \frac{d\Sigma}{dt},  
\label{dedt-gen}
\end{equation}
where $\langle \alpha \rangle$ is assumed to be independent of the time because $\alpha $ is constant along each magnetic field line so that its mean value is set by the boundary conditions at $F_{1}$ on the stellar surface that is not perturbed by the planet. For the sake of simplicity, we compute the variations of $\Theta$ and $\Sigma $ by assuming that the surface $F_{2}$ is fixed \citep[see, e.g., ][ for the neglected terms]{Smirnov64}. As ${\vec A}_{\rm p}$ and $\psi$ are completely determined by the normal component of the magnetic field $B_{\rm n}$ that does not change versus the time (see Sect.~\ref{h-variation}), their time derivatives vanish. Therefore we have:
\begin{eqnarray}
\frac{d\Theta}{dt} &  = & -  \int_{F_{2}} [( {\vec v} \times {\vec B}) \times {\vec A}_{\rm p}] \cdot \hat{\vec n} \ dS = \nonumber \\
& = &  \int_{F_{2}}   \left[ \left( {\vec A}_{\rm p} \cdot {\vec B} \right) v_{\rm n} - \left( {\vec A}_{\rm p} \cdot {\vec v} \right) B_{\rm n} \right] \ dS,
\label{dthetadt}
\end{eqnarray}
where we have made use of $\frac{\partial {\vec A}}{\partial t} = {\vec v} \times {\vec B}$ in the Coulomb gauge and $v_{\rm n}$ and $B_{\rm n}$ are the normal components of the velocity and the magnetic field on $F_{2}$. Similarly, the variation of the surface term $\Sigma$ is:
\begin{eqnarray}
\frac{d \Sigma}{d t} & = & \frac{1}{2\mu} \frac{d}{dt} \int_{F_{2}}  ({\vec A}Ê\times {\vec B}) \cdot \hat{\vec n} dS = \nonumber \\
 & = & \frac{1}{2\mu} \int_{F_{2}} \left\{ ({\vec v} \times {\vec B} ) \times {\vec B} + {\vec A} \times [\nabla \times ({\vec v} \times {\vec B} ) ]    \right\} \cdot \hat{\vec n} dS, 
\end{eqnarray}
where we have made use of the induction equation of ideal MHD and of $\frac{\partial {\vec A}}{\partial t} = {\vec v} \times {\vec B}$. The contribution of  $d\Sigma/ dt $ to the energy variation is therefore of the order of $v B^{2} F_{2}/2\mu$. Considering a maximum cross section of the flux tube equal to the cross section of the planetary magnetosphere, i.e., $F_{2} = \pi R_{\rm m}^{2}$, the contribution to the power is:
\begin{equation}
 \frac{d\Sigma}{dt} \approx  \frac{\pi}{2\mu}  B^{2} R_{\rm m}^{2} v.
\label{dsigmadt}
\end{equation} 
This is comparable with the energy released by the reconnection between the stellar and  planetary magnetic fields  on the boundary of the planetary magnetosphere (cf. Eq.~\ref{rec-power}), as discussed in Sect.~\ref{reconnection}. 

Considering this estimate of the magnitude of $d\Sigma /dt$, we see that when $2  \langle \alpha \rangle ({\vec A}_{\rm p} \cdot {\vec B}) \gg B^{2}$ the terms containing the  derivatives of the magnetic helicity  and of  $\Theta$ dominate over the derivative of  $\Sigma$ in Eq.~(\ref{dedt-gen}) and we can write:
\begin{equation}
\frac{dE}{dt} \simeq \frac{ \langle \alpha \rangle }{2\mu} \left( \frac{dH_{\rm R}}{dt} +\frac{d\Theta}{dt} \right).
\label{energy-hrel}
\end{equation}
Note that Eq.~({\ref{energy-hrel}) provides a lower limit for the dissipated power because it was obtained in the hypothesis that the field is always close to the minimum energy state corresponding to its relative helicity. If the field has a larger excess energy, i.e., it is in a non-linear force-free state far from the minimum energy state, then a greater energy can be released. 

\subsection{Variation of the magnetic helicity and energy}
\label{h-variation}

\begin{figure*}
\centering{
\includegraphics[width=16cm,height=13cm,angle=0]{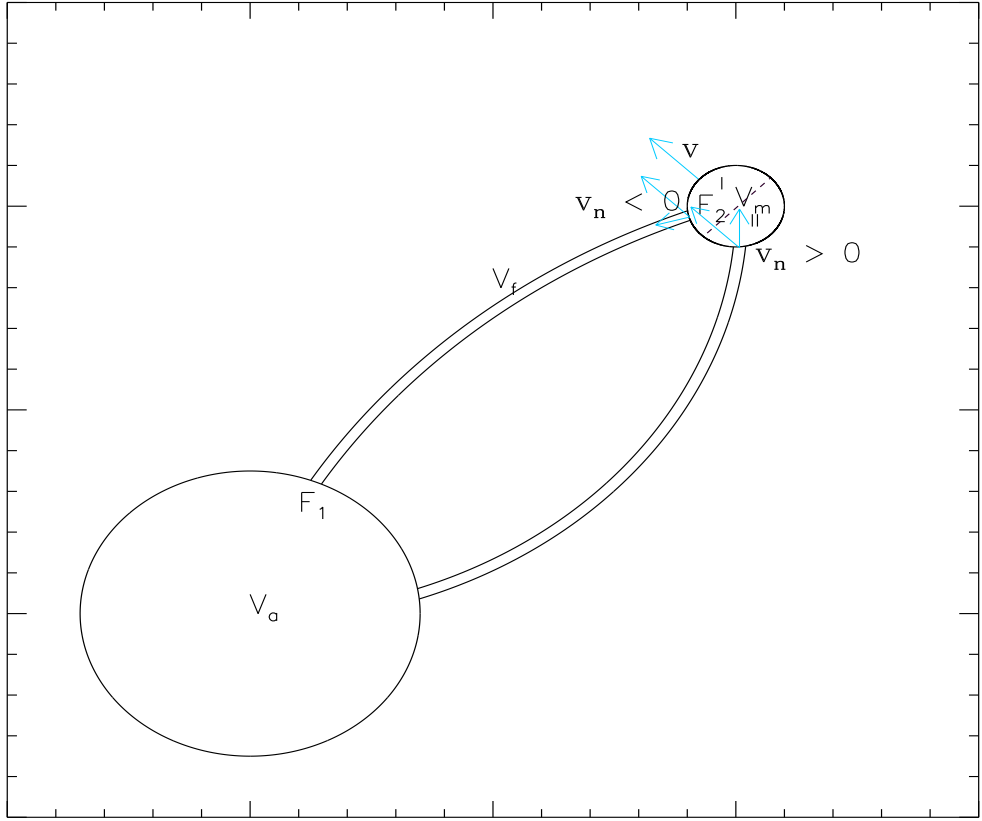}} 
\caption{A sketch of  the star-planet system with indication of  $V_{\rm a}$, the volume  interior to the star, and $V_{\rm m}$, the volume occupied by the planetary magnetosphere assumed to be spherical. Two flux tubes are plotted and in the upper one $V_{\rm f}$ the cross sections $F_{1}$ and $F_{2}$ where the tube intersects the stellar surface and the magnetosphere are labelled, respectively. 
The blue arrows indicate the relative orbital velocity $\vec v$ and its normal component $v_{\rm n}$ at the intersections of the two flux tubes with the magnetosphere. 
The dashed line inside $V_{\rm m}$ separates the region I where $v_{\rm n} < 0$, i.e., is oppositely directed with respect to the outward normal to the base of the flux tube, from the region II where $v_{\rm n} > 0$.  }
\label{magnetosphere_sketch}
\end{figure*}

The relative magnetic helicity  varies according to the following formula
derived by applying the induction equation of ideal MHD to the definition of relative helicity \citep[see ][]{BergerField84,Veksteinetal93}:
\begin{equation}
\frac{dH_{\rm R}}{dt} = -2 \oint_{S} \left[ \left( {\vec A}_{\rm p} \cdot {\vec B} \right) v_{\rm n} - \left( {\vec A}_{\rm p} \cdot {\vec v} \right) B_{\rm n} \right] \ dS,
\label{hvar}
\end{equation}
where $S$ is the boundary of the domain occupied by the field and $v_{\rm n}$ and $B_{\rm n}$ the normal components of the plasma flow $ \vec v $ and of the magnetic field $\vec B $ on $S$, and $\vec A_{\rm p}$  the vector potential of the potential magnetic field with the same normal component as $\vec B$ on $S$;  the unit normal is taken  positive in the direction outward the volume occupied by the field.  

The emergence of the magnetic flux from the convection zone of the star and the velocity fields at the base of  the corona produce a continuous variation of the relative helicity that can be associated with the energy dissipation responsible for the coronal heating and the storage of excess energy to be delivered in CMEs. These processes are independent of the presence of the planet and contribute to the unperturbed energy budget of the stellar corona. Here we  focus on the processes related to the planet. In the framework of our hypotheses, they are related to the magnetic and the velocity fields on the interface $S_{\rm m}$ between the planetary magnetosphere and the coronal field. The surface $S_{\rm m}$ is a closed surface and occupies a limited volume of the corona. For the sake of simplicity, we have assumed  it to have a spherical shape. The magnetic field is tangent to $S_{\rm m}$ because it separates the coronal field from the planetary field and the pressure of the latter forbids the penetration of the former into the planetary magnetosphere. 
Since the radius of $S_{\rm m}$ is small in comparison with the lengthscale across which the coronal field varies, we can assume that the local field at a generic point on $S_{\rm m}$ is ${\vec B}_{\rm m} = {\vec B} + {\vec B^{\prime}}$, where $\vec B$ is the  unperturbed field of the corona (assumed uniform) and $\vec B^{\prime}$  a local perturbation that makes the  local field tangent to the magnetosphere, i.e., ${\vec B}_{\rm m}\cdot \hat{\vec n}_{\rm s} = 0$, where ${\vec n}_{\rm s}$ is the  normal to $S_{\rm m}$. Since the magnetic energy is steadily dissipated by reconnection on the boundary of the magnetosphere, the energy of the field ${\vec B}_{\rm m}$ is minimized. The minimum energy of the  field ${\vec B}_{\rm m}$ compatible with the prescribed boundary condition is obtained when ${\vec B}^{\prime}$ is a potential field, i.e., ${\vec B}^{\prime} = \nabla \zeta$, where the potential $\zeta$ satisfies Laplace equation $\nabla^{2} \zeta = 0$ and ${\vec B}^{\prime}$ decreases to zero at the infinity since the effects of the magnetosphere are localized. The solutions of Laplace equation that vanish at the infinity are proportional to $1/r$ and its derivatives with respect to the spatial coordinates, where $r$ is the distance from the centre of the planet. Given the complete symmetry of a spherical magnetosphere, only the constant vector $\vec B$ can appear in the solution and, considering the linearity of both Laplace equation and the boundary conditions, $\zeta $ must involve $\vec B$ linearly. The only scalar that can be formed from $\vec B$ and the derivatives of $1/r$ is the scalar product ${\vec B} \cdot \nabla (1/r)$ \citep[cf. the mathematically analogous case of the potential flow of an ideal incompressible fluid around a spherical body moving through the fluid, e.g.,  ][]{LandauLifshitz59}. Therefore, we seek $\zeta $ in the form:
\begin{equation}
\zeta = {\vec C} \cdot \nabla \left( \frac{1}{r} \right) = - \left( {\vec C} \cdot \hat{\vec r} \right) \frac{1}{r^{2}},
\end{equation}
where $\vec C$ is a constant vector that is chosen to satisfy the boundary condition on the sphere $S_{\rm m}$, i.e., ${\vec B}^{\prime} \cdot \hat{\vec r} = - {\vec B} \cdot \hat{\vec r} $. Since $B_{r}= \partial \zeta / \partial r$, we immediately find $\vec C = -(R_{\rm m}^{3}/2) {\vec B}$ yielding: 
\begin{equation}
\zeta = \frac{1}{2} ({\vec B} \cdot \hat{r}) \frac{R_{\rm m}^{3}}{r^{2}}.
\label{zeta-express}
\end{equation}
The perturbation of the magnetic field is: 
\begin{equation}
{\vec B}^{\prime} = \nabla \zeta = \frac{1}{2} \left( \frac{r}{R_{\rm m}} \right)^{-3} \left[ {\vec B} -3 ({\vec B} \cdot \hat{\vec r}) \hat{\vec r} \right].
\label{bprime}
\end{equation}
The variation of the  magnetic energy of the coronal field produced by the perturbation ${\vec B}^{\prime}$  is (see Appendix~\ref{field-energy}):
\begin{equation}
\Delta E = - \frac{3}{4 \mu} V_{\rm m} B^{2} = - \frac{\pi}{\mu} R_{\rm p}^{3} B_{\rm p0} B,
\label{energy-field-pertur}
\end{equation}
where $V_{\rm m} = (4\pi/3) R_{\rm m}^{3}$ is the volume of the magnetosphere, $R_{\rm p}$ the radius of the planet, $B_{\rm p0}$ the field intensity at the pole of the planet, and $B$ the intensity of the coronal field at the location of the planet. Considering a magnetic loop that is crossed by the planetary magnetosphere, its energy will change by $\Delta E$ over a timescale  $\tau_{\rm cross} = 2 R_{\rm m}/v_{\rm rel}$, where $v_{\rm rel}$ is the relative velocity between the planet and the field. Therefore, the power dissipated is: $P = | \Delta E | /\tau_{\rm cross} = (1/2) (\pi/\mu) R_{\rm m}^{2} B^{2} v_{\rm rel}$, i.e., it is the same as the power dissipated by magnetic reconnection at the magnetospheric boundary for $\gamma_{\rm rec} =1/2$  (cf. Eq.~\ref{rec-power}). As we saw in Sect.~\ref{reconnection}, it is  insufficient for our purposes, thus we proceed with evaluating the energy released by the magnetic helicity variation. 

The potential magnetic field ${\vec B}_{\rm pm}$ satisfying the same boundary condition of ${\vec B}_{\rm m}$ on  $S_{\rm m}$ can be found with a similar argument. We consider its vector potential ${\vec A}_{\rm pm} = {\vec A}_{\rm p} + {\vec A}_{\rm p}^{\prime}$,  where ${\vec A}_{\rm p}$ is the vector potential of the unperturbed  potential field ${\vec B}_{\rm p}$ that depends on the boundary conditions at the stellar surface, and ${\vec A}_{\rm p}^{\prime}$ is its local perturbation. The local perturbation of the potential field ${\vec B}_{\rm p}^{\prime} = \nabla \times {\vec A}_{\rm p}^{\prime}$ can be derived by the same argument applied above to find ${\vec B}^{\prime}$ yielding ${\vec B}_{\rm p}^{\prime} = (1/2) (r/R_{\rm m})^{-3} [{\vec B}_{\rm p} - ({\vec B}_{\rm p} \cdot \hat{\vec r}) \hat{\vec r} ]$.  The corresponding vector potential in the Coulomb gauge is: 
\begin{equation}
{\vec A}_{\rm p}^{\prime} = \frac{1}{2} \left( \frac{r}{R_{\rm m}} \right)^{-3} ( {\vec r} \times {\vec B}_{\rm p} ).
\label{aprime}
\end{equation}
As in Sect.~\ref{helicityandenergy}, we consider a magnetic flux tube connecting the surface of the star with the magnetospheric boundary $S_{\rm m}$ and indicate by $F_{2}$ its base on $S_{\rm m}$ (cf. Fig.~\ref{magnetosphere_sketch}).   
The rate of change of the magnetic energy of the flux tube  produced by  the variation of the relative helicity and of the surface term $\Theta$ can be computed by Eqs.~(\ref{energy-hrel}), (\ref{hvar}), 
 and (\ref{dthetadt}) and is:
 \begin{eqnarray}
 \frac{dE}{dt} & = & - \frac{\langle \alpha \rangle}{2 \mu} \oint_{F_{2}} \left[ \left( {\vec A}_{\rm m p} \cdot {\vec B}_{\rm m} \right) v_{\rm n} - \left( {\vec A}_{\rm m p} \cdot {\vec v} \right) B_{\rm m n} \right] \ dS =\nonumber \\ 
& = &   - \frac{\langle \alpha \rangle}{2 \mu} \oint_{F_{2}}  \left[ ({\vec A}_{\rm  p} + {\vec A}_{\rm p}^{\prime}) \cdot ({\vec B}   + {\vec B}^{\prime}) \right] v_{\rm n}  \  dS,
 \end{eqnarray}
because $B_{\rm mn} = - {\vec B}_{\rm m} \cdot \hat{\vec n}_{\rm s}= 0$ on the boundary of the magnetosphere. Since ${\vec B}$ and ${\vec A}_{\rm p}$ can be considered uniform on $S_{\rm m}$, on account of Eqs.~(\ref{bprime}) and~(\ref{aprime}), we obtain:
\begin{eqnarray}
 \frac{dE}{dt} & = & - \frac{\langle \alpha \rangle}{2 \mu} \oint_{F_{2}} \frac{3}{2} \left[ ({\vec A}_{\rm p} \cdot {\vec B}) - ({\vec A}_{\rm p} \cdot \hat{\vec r}) ({\vec B} \cdot \hat{\vec r}) \right] v_{n} \ dS + \nonumber \\
   & - & \frac{\langle \alpha \rangle}{2 \mu} \oint_{F_{2}} \frac{3}{4} [({\vec r} \times {\vec B_{\rm p}} ) \cdot {\vec B}] v_{\rm n} \ dS.
   \label{eq25}
\end{eqnarray}
The second integral on the r.h.s. can be neglected in comparison to the first one because $| {\vec r} \times {\vec B}_{\rm p} | \approx R_{\rm m} B_{\rm p} $  is much smaller than $| {\vec A}_{\rm p} |$ that is of the order of $ a B_{\rm p}$, where $a$ is the semimajor axis of the planetary orbit. 

To compute the first integral, we adopt  a Cartesian reference frame with the $\hat{z}$ axis along the relative orbital velocity between the planet and the stellar field ${\vec v}$ and the $\hat{x}$ axis chosen so that the vector $\vec B$ lies in the $xz$ plane. Moreover, we choose the base $F_{2}$ of the connecting flux tube to coincide with the hemisphere of $S_{\rm m}$ where $v_{\rm n}$ has a constant, say, negative, sign (cf. Fig.~\ref{magnetosphere_sketch}). Performing the integration over that hemisphere, we find:
\begin{equation}
\frac{dE}{dt} = \frac{3 \pi  \langle \alpha \rangle}{4 \mu} R_{\rm m}^{2} v_{\rm rel} [ ({\vec A}_{\rm p} \cdot {\vec B}) +\frac{1}{4} (2 B_{\rm z} A_{\rm p z} - B_{\rm x} A_{\rm px} ) ], 
\label{dedtfromapb}
\end{equation}
where $v_{\rm rel}$ is the relative velocity between the planet and the stellar coronal field. When the stellar field is axisymmetric, the orbit of the planet is circular and lying in the equatorial plane of the star, the vector potential ${\vec A}_{\rm p} = A_{\rm p} \hat{\vec z}$ (cf. Appendix~\ref{App1}) and Eq.~(\ref{dedtfromapb}) simplifies to:
\begin{equation}
\frac{dE}{dt} = \frac{9 \pi  \langle \alpha \rangle}{8 \mu} R_{\rm m}^{2} v_{\rm rel} ({\vec A}_{\rm p} \cdot {\vec B}). 
\label{dissipation_rate}
\end{equation}
 
 \subsection{Application to the star-planet interaction}
   
Two kinds of interaction occur on the boundary of the planetary magnetosphere $S_{\rm m}$ from the point of view of the helicity variation. Specifically, a coronal flux tube whose field lines touch the magnetosphere in the domain where $v_{\rm n} < 0$, i.e., on the left of the dashed line inside $V_{\rm m}$   in Fig.~\ref{magnetosphere_sketch}, will experience an increase of its relative helicity which reduces the amount of magnetic free energy, thus opposing  dissipation. This helicity increase can  trigger a CME with an associated flare if the helicity previously accumulated is close to the threshold for the loss of equilibrium of the field configuration \citep[cf., e.g., ][]{Zhangetal06}. On the other hand, a flux tube whose field lines touch the magnetosphere in the region where $v_{\rm n} > 0$ (i.e., on the right of the dashed line in Fig.~\ref{magnetosphere_sketch}) will experience a decrease of the relative magnetic helicity leading to an increase of the magnetic free energy and an enhancement of its dissipation. In the case of a coronal loop as sketched in Fig.~\ref{magnetosphere_sketch}, the helicity increase and decrease in the two legs compensate for each other because the magnetic field  recovers the initial unperturbed configuration after the passage of the planet through the top of the loop. 
The only net effect is a modulation of the magnetic energy dissipation that is initially reduced and then increased by the passage of the planetary magnetosphere across the coronal field lines. The total energy budget of the corona is not affected by this process, thus the X-ray luminosity of the star is not affected when  averaged along one orbital period of the planet. Note, however, that the increase of helicity produced in the first part of the modulation, i.e., in the flux tubes for which $v_{n} < 0$, may trigger a CME or a flare when a previous accumulation of helicity has brought the magnetic configuration close to the threshold for instability.  This mechanism may in principle explain the modulation of the flaring activity suggested by \citet{Pillitterietal11} in the case of HD~189733. 

{ 
These predictions are valid only if  the timescale for the energy dissipation $t_{\rm R}$ is remarkably shorter than the timescale for the helicity variation induced by the planetary motion, i.e., $t_{\rm R} \ll t_{\rm H}$, otherwise the effects of the helicity increase and decrease would be averaged to zero along  the timescale $t_{\rm R}$ and no significant energy dissipation could be observed. Moreover,  if $t_{\rm R} \geq t_{\rm H}$, the phase lag between the planet and the chromospheric hot spot should continuously vary because of the statistical character of the energy dissipation that implies a range of delays between the perturbation induced by the planet and the energy release. }

In our treatment, we have considered only the velocity field arising from the orbital motion of the planet across the magnetic field of the stellar corona. 
However, other velocity fields  may potentially be relevant for the variation of the magnetic helicity, e. g., those associated with the reconnection of the magnetic field lines or the evaporation of the planetary atmosphere. The reconnected field lines must leave the region where reconnection has occurred and this produces a flow that is nearly orthogonal to the incoming flow, i.e., the flow carrying the opposite field lines into  the reconnection region. Its speed is of the order of the local Alfven speed, but in our case it is nearly tangent to the magnetospheric boundary with a small component outward from it because the magnetic pressure of the planetary field halts the penetration of the reconnected field lines into the magnetosphere. Therefore, this velocity field will lead to $v_{\rm n} < 0$ producing an increase of helicity. The same is true for the evaporation flow of the planetary atmosphere that has been observed in some systems,  notably HD~209458 and HD~189733, and is thought to be induced by the strong irradiation by the close host star \citep[e. g.,][]{Lecavelierdesetangsetal04,Ehrenreichetal08,Linskyetal10}. The speed of the evaporation flow at the limit of the magnetospheric boundary is still uncertain, but a reasonable estimate is a few tens of km~s$^{-1}$, i.e., significantly smaller than the orbital velocity of the planet \citep{Adams11}. In conclusion, the main contribution to the modulation of the magnetic helicity comes from the orbital motion. 

\subsection{Magnetic field configurations considered for the computation of the energy dissipation rate}

To compute the amplitude of the modulation of the energy dissipation rate by means of Eqs.~(\ref{dedtfromapb}) and (\ref{dissipation_rate}), we now assume a spherical polar reference frame with coordinates $(r, \theta, \phi)$, the origin at the centre of the star and the polar axis along the stellar rotation axis. Our reference frame rotates with the stellar angular velocity $\Omega$ with respect to a distant observer.  

To obtain quantitative predictions from our model, we need to specify the configuration of the stellar magnetic field. 
Linear force-free fields obey an Helmoltz equation and can be expressed analytically both in the axisymmetric and non-axisymmetric cases allowing us to study the role of axisymmetry on the SPMI. Their main limitation is  that they  consist of a set of infinite disjoint subdomains each of which is confined between two concentric spherical surfaces.
In other words, we must restrict our consideration to a field between the surface of the star and some magnetic surface at a radius $r_{\rm L}$ as in \citet{Lanza09}. This field  cannot  extend to the infinity which leads to a severe restriction imposed on the topology of the coronal field. Nevertheless, in view of their  mathematical simplicity and the possibility of treating non-axisymmetric configurations, we shall consider in detail  linear force-free fields. 
On the other hand, non-linear force-free fields pose formidable mathematical problems. Therefore, we shall limit ourselves to the simple configurations introduced by \citet{LowLou90} and then consider an extension of  our results to the more general configurations introduced by \citet{Flyeretal04}. 

\subsection{Linear force-free fields}

First, we consider  a linear force-free field as introduced by \citet{ChandrasekharKendall57}. The dependence of the vector potential ${\vec A}_{\rm p}$ on the  distance $r$ from the star is given by Eqs.~(\ref{ap-comp}) and~(\ref{phi-expr}) in Appendix~\ref{App1}. It varies as $(r/R)^{-n}$ where $n \geq 1$ is the order of the field multipole and $R$ is the radius of the star. Since $r/R \sim 7-10$ in the case of close-in planets, the leading term is that corresponding to the dipole, i.e., $n=1$, and we can neglect the terms of order $n >1$ because their relative contributions to the energy dissipation rate are of the order of $\sim (r/R)^{-2n}$.  With these assumptions and approximations, we can estimate the order of magnitude of the dissipated power by considering for simplicity only the term ${\vec A}_{\rm p} \cdot {\vec B}$ at the  location of the planet ${\vec r}_{\rm p} \equiv (r, \theta, \phi)$ in Eqs.~(\ref{dedtfromapb}) and (\ref{dissipation_rate}); we find (cf. Appendixes~\ref{App1} and \ref{field-express}):
\begin{eqnarray}
\lefteqn{({\vec A}_{\rm p} \cdot {\vec B})_{{\vec r}_{\rm p}}  = } & &   \nonumber  \\
& & \frac{1}{4} \frac{q_{0}^{2}}{\alpha} \left( \frac{q_{0}}{q}\right)^{3} g(q)  \left\{   B_{0}^{2} \sin^{2} \theta - B_{1} B_{0} \sin 2 \theta \cos (\phi - \xi)   \, +  \right. \nonumber \\
& + & \, \left. B_{1}^{2} \left[ \sin^{2} (\phi - \xi) + \cos^{2} \theta \cos^{2} (\phi - \xi) \right] \right\},
\label{ap.b}
\end{eqnarray}
where $B_{0}$ is the intensity of the axisymmetric component of the magnetic field (i.e., the mode with $m=0$) at the poles of the star, $B_{1}$ the intensity of the radial component of the non-axisymmetric  field (i.e., the mode with $m=1$) on the equatorial plane ($\theta = \frac{\pi}{2}$)  at the longitude $\xi$, $\alpha$  the force-free parameter, $q_{0} \equiv |\alpha | R$, $q= |\alpha | r$, and the function $g(q)$ has been defined in Eq.~(2) in Sect.~3 of \citet{Lanza09}. For completeness, we provide the expressions of the magnetic field components in Appendix~\ref{field-express} together with that of $g(q)$. 

Assuming that the planetary orbit is on the equatorial plane ($\theta = \pi/2$) and is circular with a radius $a$, the expression of $({\vec A}_{\rm p} \cdot {\vec B})$ simplifies to:
\begin{equation}
({\vec A}_{\rm p} \cdot {\vec B})_{{\vec r}_{\rm p}} =  \frac{1}{4} \alpha R^{2} \left( \frac{q_{0}}{q} \right)^{3} g(q) \left[ B_{0}^{2} + B_{1}^{2} \sin^{2} (\phi - \xi) \right], 
\label{ab-eq}
\end{equation}
where $q \equiv |\alpha | a $.

 The squared intensity of the stellar magnetic field $ \vec B$ at the distance of the planet is given by (cf. Appendix~\ref{field-express}):
\begin{eqnarray}
B^{2} ({\vec r}_{\rm p}) & = & \frac{1}{4} \left( \frac{q_{0}^{4}}{q^{2}} \right) \left\{ [g^{\prime}(q)]^{2} + [g(q)]^{2} \right\} \left[ B_{0}^{2} + 
B_{1}^{2} \sin^{2} (\phi - \xi) \right] + \nonumber \\
 & + & \left( \frac{q_{0}}{q} \right)^{4} [g(q)]^{2} B_{1}^{2} \cos^{2} (\phi - \xi).    
 \label{b2}
\end{eqnarray}
The magnetospheric radius $R_{\rm m}$ follows by substituting Eq.~(\ref{b2}) into Eq.~(\ref{rm}). We immediately see that $R_{\rm m}$ is not constant when $B_{1} \not= 0$, but  is modulated by the terms $\cos^{2} (\phi - \xi)$ and $\sin^{2} (\phi -\xi)$, i.e., it has a time frequency of $2(\omega -\Omega)$, where $\omega$ is the orbital frequency of the planet. 

 Eqs.~(\ref{ab-eq}) and (\ref{rm}) provide the basic ingredients to compute the energy dissipation rate from Eqs.~(\ref{dedtfromapb}) and (\ref{dissipation_rate}). It is useful to consider separately the cases of an axisymmetric coronal field and that of a non-axisymmetric field to point out the difference in the frequency of the modulation of the magnetic dissipation rate.
 
 \subsubsection{Axisymmetric linear force-free field}
 
 When $|B_{0}| \gg |B_{1}|$, the axisymmetric field dominates and the amplitude of the modulation of the energy dissipation rate as given by Eq.~(\ref{dissipation_rate}) becomes:
 \begin{equation}
P_{\rm hel}   =  
 \frac{9}{2} 4^{-\frac{7}{6}} \frac{ \pi}{ \mu} R_{\rm p}^{2} B_{0}^{\frac{4}{3}} B_{\rm p0}^{\frac{2}{3}} q_{0}^{\frac{11}{3}} q^{-\frac{7}{3}} g(q) \left\{ [g^{\prime}(q) ]^{2} + [g(q) ]^{2} \right\}^{-\frac{1}{3}} v_{\rm rel}. 
\label{axis-diss}
 \end{equation}
This expression has the same dependence on $R_{\rm p}$, $B_{0}$,  $B_{\rm p0}$, and $v_{\rm rel}$ of the energy dissipation rate produced by the reconnection between the coronal and the planetary fields (cf. Sect.~\ref{reconnection}).  
The ratio between the two powers is: 
\begin{equation}
\frac{P_{\rm hel}}{P_{\rm rec}} = \frac{9}{8 \gamma_{\rm rec}} \left( \frac{q_{0}}{q} \right) \frac{g(q)}{[g^{\prime}(q)]^{2} + [g(q)]^{2}},
\label{diss-ratio}
\end{equation}
where   $q= |\alpha | a$  is the non-dimensional orbital radius of the planet. In the distance range where close-in planets are usually found, i.e., $q/q_{0} \sim 7-10$, the factor containing the function $g$ and its derivative is of the order of the unity. Therefore, the energy released by the reconnection is generally greater than the energy released by the helicity dissipation and the derivative of the surface term $\Sigma$ in Eq.~(\ref{dedt-gen})
cannot  be neglected. 

The chromospheric emission corresponding to the energy release occurring in the corona is localized at the footprints of the flux tube that at each given time experiences a reconnection or a decrease of its helicity (cf. Sect.~\ref{h-variation}). Therefore, a distant observer will see a modulation of the chromospheric emission with the orbital period of the planet since the amplitude of the modulation is independent of the orbital or rotation phase, given the axisymmetry of the field, and  the visibility of the footprints is modulated by the motion of the planet along its orbit. This model prediction corresponds to the observations of the chromospheric hot spots in HD~179949 and $\upsilon$~And as reported by 
\citet{Shkolniketal05,Shkolniketal08}. The phase lag between the longitude of the planet and the chromospheric footprints of the flux tube interacting with the planetary field has been discussed by \citet{Lanza08} and we refer to that study for more details.

\subsubsection{Non-axisymmetric linear force-free field}
\label{non_axisfield}

When $| B_{1} |$ is comparable or larger than $| B_{0} |$, the situation is more involved. 
The power dissipated  in the interaction when $| B_{1} | \approx | B_{0}|$ is of the same order of magnitude of that released  in the case of an axisymmetric field, but it depends explicitly on the time because $ ({\vec A	}_{\rm p} \cdot {\vec B})$, $2B_{\rm z} A_{\rm pz} - B_{\rm x}A_{\rm px}$, and  $R_{\rm m}$ are modulated by the terms $\sin^{2} (\phi -\xi )$ and $\cos^{2} (\phi -\xi )$
(cf. Eqs.~\ref{rm}, \ref{ab-eq}, and \ref{b2}).  In other words, $P_{\rm rec}$ and $P_{\rm hel}$ are modulated with the frequency $ 2(\omega -\Omega)$, i.e., twice the synodic frequency. From a rigorous point of view, the magnetospheric radius $R_{\rm m}$ is a   function of $B^{2}$ which introduces  contributions from the higher harmonics, i.e., $4 (\omega -\Omega)$, $6 (\omega -\Omega)$, etc.,  into the modulation of the energy dissipation rates. For simplicity, we assume that the fundamental frequency, i.e., $2(\omega -\Omega)$, dominates the variation of $R_{\rm m}$, leading to the same frequency 
for the modulation of $P_{\rm rec}$ and $P_{\rm hel}$. 
Moreover, one should consider that the visibility of the chromospheric footprints  in the case of a non-axisymmetric field is modulated by the rotation of the star which adds a further characteristic frequency, i.e. $\Omega$, and its harmonics to our variations. Typically, the first two harmonics, i.e., $2\Omega$ and $3 \Omega$ should be considered. 

The combination of the modulation of $P_{\rm rec}$ and $P_{\rm hel}$ with the frequency $2(\omega-\Omega)$ and the modulation of the visibility of the footprints with the frequency $\Omega$ and its harmonics produces  a  time dependence of the signal with  frequencies $(2 \omega -\Omega)$,  $(2\omega - 3\Omega)$, $2 \omega$, $2\omega -4 \Omega$, $2\omega - 5\Omega$, and $2 \omega + \Omega$  for a distant observer.  Thus, the modulation of the chromospheric emission with the orbital frequency is no longer the dominant periodicity. This may  lead to a confused situation in which it is virtually impossible to disentangle the SPMI signal from the intrinsic variability and the rotational modulation of the chromospheric emission typical of an active star \citep[see also the study by][]{CranmerSaar07}. This  can explain why in several cases there has been no evidence of SPMI, even in those systems that have shown a modulation of the chromospheric emission with the orbital period in some seasons, such as HD~179949 or $\upsilon$~And. Considering the Sun as a template, we see that the large scale magnetic field is close to be axisymmetric during the minima of the eleven-year cycle, while it shows a  deviation from axisymmetry during the other phases of the cycle when  large active regions or complexes of activity are present on the surface. 

The prediction that the modulation of the SPMI signal becomes practically indistinguishable from intrinsic stellar activity variations when the photospheric field has a sizable non-axisymmetric component can be tested by applying Zeeman Doppler Imaging techniques along the lines of the studies  by, e.g., \citet{Moutouetal07} and \citet{Faresetal10}. 

\subsection{Non-linear fields}
\label{non-linear-field}

In view of the high mathematical complexity of non-linear force-free fields, we limit ourselves to some specific kinds of  axisymmetric fields that can be expressed as:
\begin{equation}
\vec B = \frac{B_{0} R^{2}}{r \sin \theta} \left[  \frac{1}{r} \frac{\partial A}{\partial \theta} \hat{\vec r} - \frac{\partial A}{\partial r} \hat{\vec \theta} + \frac{1}{R} Q(A)  \hat{\vec \phi} \right],
\label{lowloufield}
\end{equation}
where $B_{0}$ sets the radial field intensity at the North pole, $A(r, \theta)$ is the flux function, and $Q=Q(A)$  a scalar function that has a different functional form according to the specifically considered family of fields. Note that both $A$ and $Q$  are non-dimensional in our definition. The families of fields considered by \citet{LowLou90} have a separable flux function of the form:
\begin{equation}
A(r, \theta) = (r/R)^{-n} f(x),
\label{wolfsondecay}
\end{equation}
where $x \equiv \cos \theta$,  $n$ is a positive constant, not necessarily an integer, and $f$ is given by the differential equation:
\begin{equation}
(1 -x^{2}) f^{\prime \prime}(x) + n (n+1) f(x) + \lambda^{2} \left( 1 + \frac{1}{n} \right) [f(x)]^{1+ 2/n} = 0,
\end{equation}
that is solved in $[-1, 1]$ subject to the boundary conditions $f(-1)=f(1)=0$ with $\lambda^{2}$ as an eigenvalue. 
The scalar function $Q$ in this case is given by:
\begin{equation}
Q(A) = \lambda A^{1 +1/n}, 
\end{equation}
while the force-free parameter $\alpha$ is:
\begin{equation}
\alpha= \frac{1}{R} \frac{dQ}{dA} = \frac{\lambda}{R} \frac{n+1}{n} \left( \frac{r}{R} \right)^{-1} [f(x)]^{\frac{1}{n}}. 
\label{alpha-def}
\end{equation}
We shall consider fields with $0 <n < 1$ because they have the slowest decay  with the distance (cf. Eq.~\ref{wolfsondecay}) and look for solutions of the boundary value problem for $f(x)$ 
that satisfy the conditions $f(-x) = f(x)$ in $[-1, 1]$ and $f^{\prime}(0) = 0$, following the method described by \citet{Wolfson95}. The magnetic field components are:
\begin{eqnarray}
B_{r} & = &  -B_{0} (r/R)^{-(n+2)} f^{\prime} (x), \nonumber \\
B_{\theta} & = & n B_{0}  (r/R)^{-(n+2)} \frac{f(x)}{\sin \theta} \\
B_{\phi} & = & \lambda B_{0}  (r/R)^{-(n+2)} \frac{[f(x)]^{1+ 1/n}}{\sin \theta}. \nonumber
\end{eqnarray}
The potential magnetic field with the same normal component $B_{\rm p r}= - B_{0} f^{\prime}(x)$ on the surface of the star is discussed in Sect.~5 of \citet{Wolfson95}. We consider only the component of its vector potential with the slowest decay with the distance from the star, i.e., that corresponding to the dipole component because it dominates over the higher order multipoles given that the multipole of order $k$ decays as $(r/R)^{-k}$. Therefore, we find ${\vec A}_{\rm p} = (3/4) B_{0} R (r/R)^{-2} \sin \theta \, \hat{\vec \phi}$ and assuming  the planet in the equatorial plane:
\begin{equation}
({\vec A}_{\rm p} \cdot {\vec B})_{{\vec r}_{\rm p}} = \frac{3}{4} B_{0}^{2} R Q(A) \left( \frac{r}{R} \right)^{-3}.
\end{equation}
To compute the energy dissipation rate from Eq.~(\ref{dissipation_rate}), we consider  the magnetospheric radius as given by Eq.~(\ref{rm}) and assume that the flux tube connecting the stellar surface with the boundary of the planetary magnetosphere has a nearly uniform force-free parameter, i.e., $\langle \alpha \rangle \simeq \alpha$, thus obtaining:
\begin{equation}
 P_{\rm hel} = \frac{27\pi}{16\mu} \frac{n}{n+1} (\alpha R)^{2} B_{0}^{4/3}B_{\rm p0}^{2/3} R_{\rm p}^{2} (\lambda^{2} + n^{2})^{-1/3}   v_{\rm rel} \left( \frac{r}{R} \right)^{-(n+5)/3}.
\end{equation}
Since $\alpha $ is constant along magnetic field lines and is given by Eq.~(\ref{alpha-def}), we  recast this equation as:
\begin{equation}
P_{\rm hel}   =   \frac{27}{16} \frac{\pi}{\mu} \frac{n+1}{n} \lambda^{2} B_{0}^{4/3}B_{\rm p0}^{2/3} R_{\rm p}^{2} (\lambda^{2} + n^{2})^{-1/3}  v_{\rm rel} \left( \frac{r}{R} \right)^{-(n+11)/3}. 
\label{en_diss-nl}
\end{equation}
{ These equations are valid provided that $\alpha ({\vec A_{\rm p}} \cdot {\vec B}) = \alpha A_{\rm p} B_{\phi}$ is  uniform  over the surface of the planetary magnetosphere (cf. Sect.~\ref{h-variation}). Since the relative variation of  $\alpha B_{\phi}$ in the meridional direction increases with decreasing $n$, Eq.~(\ref{en_diss-nl}) is valid only when $n \ga 0.1$. For smaller values of $n$ the energy dissipation rate is reduced by a factor $\approx  n r / 2 R_{\rm m}$, where $r$ is the distance of the planet and $R_{\rm m}$ the radius of its magnetosphere, as can be derived from Eq.~(\ref{eq25}). Therefore, the energy dissipation rate does not diverge for $n \rightarrow 0$, but tends to a finite limit because the eigenvalue $\lambda^{2} \rightarrow 1$ in the same limit. The maximum value of the energy dissipation is obtained  from Eq.~(\ref{en_diss-nl}) for $ n \approx 0.1$. 

The magnetic field configurations corresponding to our model have been studied in detail by \citet{Wolfson95}. When the value of $n$ decreases, the projections of the field lines on the meridional plane form taller and taller  loops. In the limit $ n \rightarrow 0$, the field lines become radially directed and an infinitely thin current sheet is formed on the equatorial plane. Such a configuration cannot support a confined corona because the hot plasma is free to escape along the radial field lines. For the existence of a confined corona, as indicated by the X-ray observations of planet-hosting stars, we should limit to values of $n > 0.1$. In this case, the total energy of the field is well below the so-called Aly limit, i.e., the minimum energy for the opening of all the field lines for the assumed boundary conditions \citep[see  Sect.~4 in ][ for more details]{Wolfson95}, and the pressure of the plasma cannot produce a loss of confinement of the corona by opening up its field lines. 
}

The ratio of the power released by the helicity modulation to that dissipated by the magnetic reconnection is:
\begin{equation}
\frac{P_{\rm hel}}{P_{\rm rec} } = \frac{27}{16} \frac{1}{\gamma_{\rm rec}} \frac{n+1}{n} \frac{\lambda^{2}}{\lambda^{2} + n^{2}} \left( \frac{r}{R} \right)^{n-1},
\end{equation}
For a typical relative distance of the close-in planets $(r/R) \sim 7-10$ and $n \sim 0.1$ the energy released by the helicity modulation is $\approx 4-5$ times larger than that produced by the reconnection between the magnetic fields of the star and the planet. This implies that Eq.~(\ref{energy-hrel}) is approximately valid for the considered magnetic fields and can be used to estimate  the energy dissipation rate neglecting the contribution of the time derivative of the surface term $\Sigma$. 

The footpoints of the field line connecting the stellar surface with the planet  are located at  colatitudes $\theta_{0}$ and 
$\pi-\theta_{0}$,  symmetric with respect to the equator. The value of $\theta_{0}$ follows from the constancy of $\alpha$ along each field line and is  the solution of the equation:
\begin{equation}
f(\cos \theta_{0}) = (r/R)^{-n},
\label{footpoint_theta}
\end{equation}
since on the equatorial plane where the planet is located $f(0)=1$. The azimuthal angle $\Delta \phi $ between the footpoints and the planet is given by Eq.~(8) of \citet{Wolfson95} that we reproduce here:
\begin{equation}
\Delta \phi = \frac{\lambda}{n} \int_{0}^{\cos \theta_{0}} \frac{[f(x)]^{1/n}}{(1-x^2)} \ dx.
\label{footpoint_phi}
\end{equation}

Another class of axisymmetric force-free fields of the kind specified by Eq.~(\ref{lowloufield}) has been proposed by \citet{Flyeretal04}. They have:
\begin{equation}
[Q(A)]^{2} = \frac{2 \gamma}{k+1} A^{k+1},
\end{equation}
where $\gamma$ is a parameter (constant for a given field configuration) and $k$  an odd positive integer which ensures that the r.h.s. of this equation is positive definite independent of the sign of the flux function $A$. The force-free equation is solved subject to dipole-like boundary conditions, i.e., $A(R, \theta) = \sin^{2} \theta$, $|\nabla A | \rightarrow 0$ as $r \rightarrow \infty$, and $A(r, \theta = 0 \mbox{ and } \pi) = 0$. With those boundary conditions, solutions to the force-free equation are found numerically only for $k \geq 5$ \citep[see ][]{Flyeretal04}. { Fields with $k > 9$ are very difficult to treat numerically owing to the steep gradients of $Q$. 

An interesting property of these fields is the possibility of an approximate asymptotic representation in the limit  $ r/R \gg 1$ by means of the Low \& Lou's fields \citep[cf. \S~2.5 of ][]{Flyeretal04}. For $k=5$, the representation by a field with $n=1/2$  is remarkably accurate, while for $k =7$ and $k=9$, the corresponding  fields with $n=1/3$ and $n=1/4$, respectively, give less good representations in the range $r/R \sim 7-10$  of interest for our model.  
}

\section{Results}
\label{results}
\subsection{Coronal parameters and timescales}
\label{coronal_param}

{
We shall consider the systems for which some evidence of SPMI has been reported, notably HD~179949, $\upsilon $~And, $\tau$~Boo, and HD~189733. 
The stellar and planetary parameters adopted in our computations are reported in Table~\ref{parameters} where we list from the left to the right, the name of the system, the surface magnetic field intensity, the rotation period of the star, the orbital period, the semimajor axis of the orbit, the ratio $a/R$, and the relative velocity $v_{\rm rel}$ between the planet and the stellar coronal field. 

The magnetic fields  of the host stars can be measured using Zeeman Doppler Imaging techniques. For the F-type stars HD~179949, $\upsilon$~And, and $\tau$~Boo, we assume $B_{0} = 10$~G which is  the maximum radial  field  observed  in $\tau $~Boo close to the visible pole  \citep{Catalaetal07,Donatietal08,Faresetal09}. For the K-type star HD~189733, we assume $B_{0}  =40$~G,  as measured by \citet{Moutouetal07} and \citet{Faresetal10}, the star being remarkably more active than the other targets. The rotation periods of HD~179949, $\upsilon$~And, and HD~189733 are taken from \citet{Shkolniketal08}, \citet{Poppenhaegeretal11}, and \citet{HenryWinn08}, respectively.  \citet{Catalaetal07} found that $\tau$~Boo has a remarkable differential rotation ($\approx 20$ percent) with a mean rotation period practically synchronized with the orbital motion of its planet. We assume a rotation period of $3.9$ days that corresponds to the high-latitude rotation period to maximize the relative velocity between the coronal field and the orbital motion of the planet.

The X-ray coronal  emissions of six planet hosting stars within 30 pc of the Sun have been fitted by \citet{Poppenhaegeretal10} with a two-temperature model giving a cool component with a temperature of about 1~MK  that dominates the emission measure. Only two stars show a comparable contribution from a hotter component with $T \sim 4-5$~MK. For HD~179949, \citet{Saaretal08} find a coronal temperature ranging from $\sim 0.45$ and $\sim 1.1$~MK with the lower temperature component having an emission measure $\approx 3$ times larger than that of the higher temperature component. In $\upsilon$ ~And, \citet{Poppenhaegeretal11} find a corona with a mean temperature of $\sim 3 $~MK. 

We apply the coronal model of Sect.~\ref{coronal_model} to estimate the plasma $\beta$ at the distance of the planets and the timescales characteristic of our systems in order to justify our approach. The variation of the magnetic field strength with the radial distance is comparable with that of a dipole field or is slower for our field configurations \citep{Lanza09}, thus we assume  $B(r) = B_{0} (r/R)^{-3}$. Indicating with $\beta_{0}$ the value of the parameter at the base of the corona, we find:
\begin{equation}
\beta(r) \equiv \frac{2 \mu p(r)}{B^{2}(r)} = \beta_{0} \left( \frac{r}{R} \right)^{6} \exp \left[ - \left( \frac{R}{H_{0}}\right) \left( 1 - \frac{R}{r}\right) \right], 
\end{equation}
where Eq.~(\ref{pressure_run}) has been applied.  Assuming an electron density  $n_{\rm e0} = 10^{14}$~m$^{-3}$ and a magnetic field of $10$~G at the base of the corona, we find $\beta_{0} = 0.007, 0.014, 0.027$ for $T=1, 2, 4$~MK, respectively. Considering a star with the same mass and radius of the Sun, and a planetary distance $r/R = 10$, we find that $\beta(r) = 10^{-4}, 0.22, 13.8$ for $T= 1, 2, 4$~MK, respectively. Therefore, we see that in a magnetostatic model the high temperature component of a stellar corona  ($T \ga 2-3$~MK) cannot extend up to the distance of the planet because there $\beta(r) \geq 1$. In other words,  the hot component of the corona is confined into loops with a height of a few  stellar radii,  while the loops extending up to the distance of the planet have a plasma temperature not exceeding $1-2$~MK. 

The other crucial assumption of our model is that the timescale for the helicity variation $t_{\rm H}$ is substantially longer than the relaxation time $t_{\rm R}$, that in turn is a few times the Alfven transit time $t_{\rm A}$ along the magnetic field lines. The timescale $t_{\rm H}$ is the time that the orbiting planet takes to travel across a distance equal to the diameter of its magnetosphere, i.e., 
$t_{\rm H} = 2R_{\rm m}/v_{\rm rel}$, where $v_{\rm rel}$ is listed in Table~\ref{parameters} for our systems. Assuming a typical magnetospheric radius $ \geq 5 \times 10^{8}$~m, we find  $t_{\rm H}$ of the order of $10^{4}$~s or longer for our systems. On the other hand, neglecting the curvature of the field lines, the Alfven transit time  is approximately given by:
\begin{equation}
t_{\rm A} \simeq \int_{R}^{a} \frac{dr}{v_{\rm A}(r)},
\end{equation}
where $v_{\rm A}(r) = B(r)/\sqrt{\mu \rho(r)}$ is the Alfven velocity at the distance $r$. Assuming that $B(r) = B_{0} (r/R)^{-3}$ and adopting the model in Sect.~\ref{coronal_model} to compute the density $\rho(r)$ with a base electron density  $n_{\rm e0}= 10^{14}$~m$^{-3}$, $B_{0}=10$~G,  $T=1$~MK, and $a/R$ as listed in Table~\ref{parameters}, we find $t_{\rm A} \sim  10^{3}$~s for HD~179949 and $\tau$~Boo; $t_{\rm A} \sim 2.2 \times 10^{3}$~s for $\upsilon$~And; and  $t_{\rm A} \sim 400$~s for  HD~189733 because  $B_{0} = 40$~G in that case. 
{Note that the orbital velocity of HD~189733 exceeds the isothermal sound speed for a coronal temperature $ \leq 10^{6}$~K that can produce a hydrodynamic bow shock in front of the planetary magnetosphere according to \citet{Vidottoetal10b,Vidottoetal11}. The shock is weak because the Mach number is $\sim 1.1-1.2$ so the magnetic field compression is negligible for our purposes \citep[see, e.g., \S~5.2.3 of ][ for the effect of a strong perpendicular shock on the magnetic field]{Vidottoetal10}. Moreover, in the assumed low-beta environment of the  stellar corona, the formation of a perpendicular shock requires that the relative orbital velocity of the planet exceeds the fast Alfven speed, i.e., $ \sqrt{2} \beta^{-1/2} c_{\rm s}$, where $c_{\rm s}$ is the  isothermal sound speed, that is not verified in our case. }

In conclusion, the requirements on the timescales characterizing the processes considered in our model are generally satisfied, provided that the temperature along the coronal field lines that  touch the planetary magnetosphere is  $\sim 1$~MK or lower, as required by the previous considerations on the confinement of the stellar corona. 

} 
\begin{table}
\caption{Stellar and planetary parameters}
\label{parameters}
\centering
\begin{tabular}{cccccccc}
\hline 
\hline
Name  & $B_{0}$ & $P_{\rm rot}$ & $P_{\rm orb}$ & $a$ & $a/R$  & $v_{\rm rel}$ \\
 & (G) & (d) & (d) & (AU) &  & (km s$^{-1}$) \\
\hline
HD179949 & 10 & 7.0 & 3.09 & 0.045 & 7.72  & 88.4 \\ 
HD189733 & 40 & 11.9 & 2.22 & 0.031 & 8.56 & 125.4 \\ 
$\upsilon$ And & 10 & 9.5 & 4.62 & 0.059 & 10.16 & 71.4 \\
$\tau$ Boo & 10 & 3.9 & 3.31 & 0.046 & 7.38 &   22.8 \\ 
\hline
\end{tabular}
\end{table}

\subsection{Linear force-free fields}
\label{linear-model-res}

We compute the energy dissipation rates for the systems considered in Sect.~\ref{coronal_param}. Considering first an axisymmetric linear force-free magnetic field, we  estimate the energy dissipation rates $P_{\rm rec}$ and $P_{\rm hel}$  from  Eq. (\ref{rec-power}) and Eq.~(\ref{axis-diss}), respectively, and compare the phase lag between the planet and the chromospheric hot spot computed by the method of \citet{Lanza08} with the observations. Since we are interested in finding the maximum released power, we increase the value of $\alpha$ as much as possible  because $P_{\rm hel}$ scales as  $q_{0}^{11/3} q^{-7/3}$ that is $|\alpha |^{4/3}$ since 
$q_{0} = | \alpha | R$ and $q = | \alpha | a$. However, a limitation on $\alpha $ is  imposed by the requirement that the semimajor axis of the planetary orbit $a$ be smaller than the limit radius $r_{\rm L} = q_{\rm L}/ |\alpha |$ where $g(q)$ has its first zero and the field lines close back onto the star. This implies in our cases $| \alpha | \la 0.5 R^{-1}$. 

To compute the radius of the magnetosphere, for the non-transiting planets of HD~179949, $\upsilon$~And, and $\tau$~Boo, we assume a radius  $R_{\rm p} = 8.6 \times 10^{7}$~m, i.e., $1.2$ Jupiter radii, while for the transiting planet of HD~189733 we have a measured radius $R_{\rm p} = 8.08 \times 10^{7}$~m. The intensity of the planetary magnetic field at the poles is not known. Here, we adopt a  field as strong as possible, i.e.,  $B_{\rm p} = 100$~G, which is about seven times the Jupiter's field, because $P_{\rm rec}$ and $P_{\rm hel}$ scale as $B_{p0}^{2/3}$ and we aim at maximizing the energy dissipation rate. Such extreme fields cannot be excluded in the framework of the dynamo models proposed by \citet{Christensenetal09} and \citet{ReinersChristensen10} provided that the internal heat flux of the planet, which controls the strength of the field, is sufficiently high. Since the dissipated power scales as $B_{\rm p0}^{2/3}$, adopting a field of $15$~G, close to the value of Jupiter, will reduce the estimated powers by a factor of $\sim 3.7$. 

We choose the parameter $b_{0}/c_{0}$ defining the function $g(q)$  to maximize the energy dissipation rate.
The parameters of our field models are reported in Table~\ref{model_field} together with the energy dissipation rates. We list from the left to the right, the name of the system, the intensity of the stellar field $B_{0}$ (see Sect.~\ref{coronal_param} for a justification of the chosen values), the intensity of the assumed planetary field $B_{\rm p0}$, the force-free parameter $\alpha$, the parameter $b_{0}/c_{0}$, the footpoint colatitude $\theta_{0}$ of the field line connecting the planet with the surface of the star, the phase lag $\Delta \phi$ between the footpoints of this line and the planet, the dissipated power $P_{\rm hel}$ as given by Eq.~(\ref{axis-diss}), and the dissipated power $P_{\rm rec}$ as given by Eq. ~(\ref{rec-power}).
\begin{table*}
\caption{Linear axisymmetric force-free fields}
\label{model_field}
\centering
\begin{tabular}{ccccccccc}
\hline
\hline
Name & $B_{0}$ & $B_{\rm p0}$ & $| \alpha | $ & $b_{0}/c_{0}$ & $\theta_{0}$ & $\Delta \phi$ & $P_{\rm hel}$ & $ P_{\rm rec}$ \\
 & (G) & (G) & ($R^{-1}$) & & (deg) & (deg) & ($10^{20}$ W) & ($10^{20}$~W) \\
 \hline
 HD 179949 & 10 & 100 & 0.5 & $-0.3$ & 40.0 & 129.23 & 0.09 & 0.61 \\
 HD 189733 & 40 & 100 & 0.5 & $-0.4$ & 20.0 & 110.99 & 0.68 & 3.10 \\
 $\upsilon$ And & 10 & 100 & 0.2 & $-1.1$ & 32.5 & 177.1 & 0.05 & 0.01\\
 $\tau$ Boo & 10 & 100 & 0.5 & $-0.5$ & 32.0 & 120.0 & 0.05 & 0.10\\
\hline
\end{tabular}
\end{table*}
\begin{figure}
\centering{
\includegraphics[width=8cm,height=12cm]{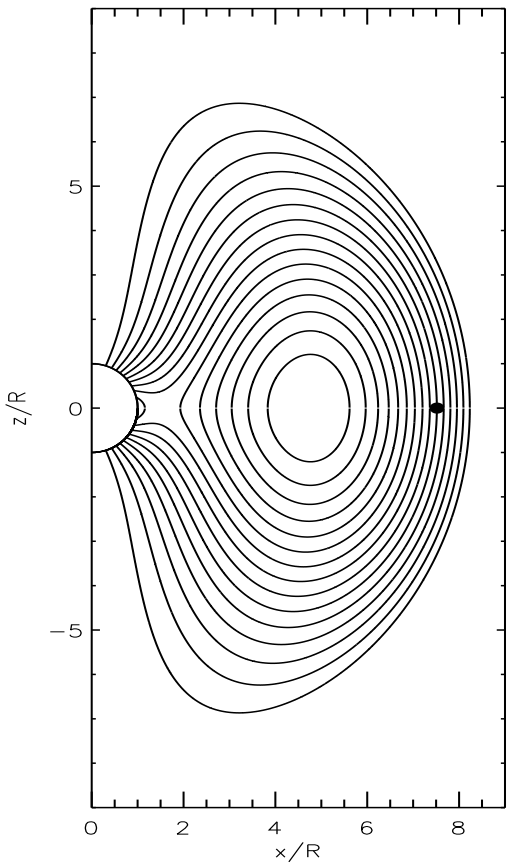}} 
\caption{Meridional section of the linear force-free magnetic field corresponding to the parameters assumed for HD~179949 in Table~\ref{model_field}. Note the rope of azimuthal flux symmetric with respect to the equator and the position of the planet indicated by the filled dot  on the equatorial plane $(z=0)$. The field line connecting the planet to the stellar surface is outside the flux rope while the field lines of the rope close onto themselves without reaching the stellar surface. }
\label{sketch-linear-field}
\end{figure}
A meridional section of the magnetic field in the case of HD~179949 is plotted in Fig.~\ref{sketch-linear-field}. The field has a prominent azimuthal flux rope symmetric with respect to the equatorial plane whose field lines are detached from the stellar surface. This kind of configuration is discussed in \citet{Lanza09}. Note that the planet is located outside  the flux rope, thus it is magnetically connected with the stellar surface and can induce the formation of a chromospheric hot spot by releasing energy in the connecting flux tube. 

The dissipated  powers listed in Table~\ref{model_field} can be compared with the observations of HD~179949 and $\upsilon$~And that give powers of $\approx 10^{20}$ and $\approx (2-3) \times 10^{19}$~W, respectively \citep{Shkolniketal05}, i.e.,  about two orders of magnitude larger than $P_{\rm hel}$ and one order of magnitude larger than $P_{\rm rec}$ as predicted by our model.  Since the dissipated power scales as $B_{0}^{4/3} B_{\rm p0}^{2/3}$, to get $P_{\rm rec}$ of the order of $10^{20}$~W, we should have $B_{0} \sim 30-50$~G, which is indeed the case for HD~189733 which is therefore the system with the highest predicted effect. The reason why there have been no conclusive observations of SPMI for this system may be  its complex magnetic field topology  with a predominance of  non-axisymmetric components \citep[cf. ][]{Moutouetal07,Faresetal10} which lead to a complicated time dependence of the dissipated power (cf. Sect.~\ref{non_axisfield}). However, the possibility of a flaring activity modulated with the orbit of the planet in HD~189733 may indicate a remarkable SPMI in this system. 

Note that $P_{\rm hel} < P_{\rm rec}$ in all of our linear force-free models. This indicates that the neglect of the surface term $d\Sigma/ dt$ in Eq.~(\ref{dedt-gen}), rigorously speaking, is not justified. The contribution of this term, however, is of the same order of magnitude of the power dissipated by the reconnection of the planetary and stellar fields, thus its inclusion is not expected to change our conclusions. 

An important constraint on the magnetic field model is provided by the lag between the planet and the phase of maximum visibility of the chromospheric hot spot. \citet{Shkolniketal05} found that the maximum chromospheric emission falls at phase $\sim 0.7-0.8$ in the case of HD~179949 and at phase$\sim 0.5$ in the case of $\upsilon$~And, where phase 0 corresponds to the inferior conjunction of the planet. In the case of $\tau $ Boo, \citet{Walkeretal08} found a phase of maximum activity around $\sim 0.7-0.8$, while for HD~189733 \citet{Shkolniketal08} suggest a possible enhancement of the intra-night chromospheric variability with a phase lag of $\approx 0.8$. 
These values correspond to angles of $\sim 80^{\circ}-120^{\circ}$ for HD~179949, $\tau$~Boo, and HD~189733, and of $\sim 180^{\circ}$ in the case of $\upsilon$~And. The $\Delta \phi$ values derived from our models with $ | \alpha | = 0.5R^{-1}$ are at the upper bounds  of those ranges for the first three stars, but are incompatible with the lag observed in $\upsilon$~And.
For this star, we cannot increase $| \alpha |$ beyond $0.2R^{-1}$, otherwise it becomes impossible to reproduce the observed phase lag. As a consequence of the smaller $| \alpha |$, we have  a remarkable reduction of the dissipated powers. 

\subsection{Non-linear force-free fields}

{  
 Here we limit ourselves to the axisymmetric fields discussed by \citet{Wolfson95}. A meridional section of the magnetic field lines in the case of HD~179949 is plotted in Fig.~\ref{wolfson-field} for $n=0.5$. 
The eigenvalue corresponding to $n=0.5$ is $\lambda^{2} = 0.82343$.
Note that this case gives also a good approximation to the more complex non-linear field of \citet{Flyeretal04} with $k=5$ in the radial range of interest for star-planet interaction. 
 In Table~\ref{table-wolfson} we list, from the left to the right, the name of the system, the colatitude $\theta_{0}$ of the footpoint of the field line joining the stellar surface with the planet as given by Eq.~(\ref{footpoint_theta}), the azimuthal angle $\Delta \phi$ between the footpoint and the planet as given by 
 Eq.~(\ref{footpoint_phi}), and the energy dissipation rates $P_{\rm hel}$ and $P_{\rm rec}$ as derived from Eq.~(\ref{en_diss-nl}) and~(\ref{rec-power}), respectively, with $B_{0} =10$~G in all the cases with the exception of HD~189733 for which $B_{0} = 40$~G. { The planetary field strength  $B_{\rm p0}= 100$~G in all the cases to maximize the energy dissipation rate as in the case of linear force-free fields. Assuming $B_{\rm p0}=15$~G -- a much more realistic value in view of the models of \citet{ReinersChristensen10} for planets with ages of at least $1-2$ Gyrs, and the field observed in Jupiter -- the given powers are reduced by a factor of $\sim 3.7$.  
 The variation of $\alpha B_{\phi}$ across the planetary magnetosphere does not exceed 5~percent in all the cases, thus our assumption of a constant coronal field over the volume of the planetary magnetosphere is well justified. 
\begin{figure}
\centering{
\includegraphics[width=8cm,height=12cm]{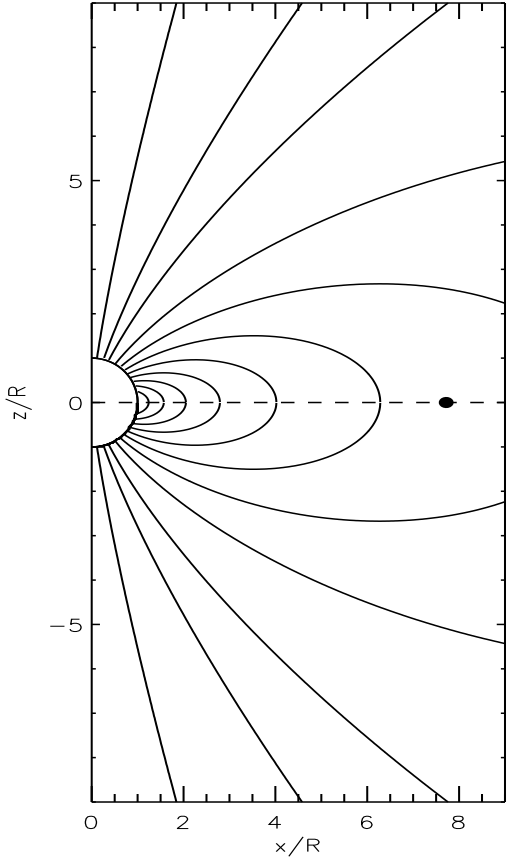}} 
\caption{Meridional section of the non-linear force-free magnetic field of \citet{Wolfson95} with $n=0.5$ for the case of HD~179949.  The filled dot indicates the close-in planet assumed to be on the equatorial plane of the star $(z=0)$ that is marked by the dashed line. }
\label{wolfson-field}
\end{figure}
\begin{table}
\caption{Non-linear axisymmetric force-free field for $n=0.5$}
\label{table-wolfson}
\centering
\begin{tabular}{ccccc}
\hline 
\hline
Name  & $\theta_{0}$ & $\Delta \phi$ & $P_{\rm hel}$ & $P_{\rm rec}$ \\
 & (deg) & (deg) & ($10^{19}$~W) & ($10^{19}$~W) \\
\hline
HD~179949 & 40.13 & 53.99 & 1.23 & 0.44 \\ 
HD~189733 & 38.66 & 54.51 & 1.04 & 0.39 \\ 
$\upsilon$~And & 36.28 & 55.27 & 0.35 & 0.14 \\
$\tau$~Boo & 40.78 & 53.75 & 0.38  & 0.13 \\ 
\hline
\end{tabular}
\end{table}

For  $n=0.25$ ($\lambda^{2}= 1.01203$),  $P_{\rm hel}$ is a factor of $2.4$ larger, but is still insufficient to account for the observations, while the ratio $P_{\rm hel}/P_{\rm rec}$ ranges from 2.8 to 3.5. 
However, $n=0.25$ gives a better agreement with the observed phase lags, because in all the cases $\Delta \phi$ ranges between $68\fdg1$ and $69\fdg7$ with  
$\theta_{0}$ between $57\fdg9$ and $60\fdg7$.  
We can decrease $n$ up to, say,  $n=0.1$, because for this value the variation of $\alpha B_{\phi}$ across the planetary magnetosphere reaches $\sim 60$~percent making our formula only roughly valid. However, even in this extreme case,  $P_{\rm hel}$ increases by a factor of  4.4 with respect to the case with $n=0.5$ and this is in the best case only marginally compatible with the observations. 

We conclude that the non-linear force-free field of \citet{Wolfson95} cannot account for the dissipated powers observed in our systems, even for extreme values of the planetary magnetic field. For  HD~179949, $\tau$~Boo, and HD~189733, it can account for the phase lag between the planet and the chromospheric hot spot if $n=0.25$, but in the case of $\upsilon$~And the predicted lag is too small.  An insufficient dissipated power is expected also in the case of the more general non-linear models  of \citet{Flyeretal04} because the Low \& Lou models give a fairly good approximation to them in the radial range considered for the star-planet interaction. 
}
}

\section{Discussion and conclusions}
\label{conclusion}

We have introduced an approximate model to compute the energy dissipated in the interaction between the magnetic field of a star and a close-in planet.  
{Our model assumes that the orbital velocity of the planet is much smaller than the Alfven speed in the stellar corona and that the coronal field perturbed by the motion of the planet relaxes to its minimum energy state  within a timescale comparable with the Alfven transit time along the coronal field lines. With these  assumptions, we can treat the evolution of the field as a sequence of magnetostatic configurations and estimate the energy variations under the constraint that the total magnetic helicity of the field is conserved. }

We have estimated the  power dissipated by magnetic reconnection as well as that released by the modulation of the field helicity  associated with the orbital motion of the planet. The latter process can operate also when the planetary field is very small or even absent because what is needed is a relative velocity  between the planet and the coronal  field of the star. If the planetary field $B_{\rm p0}=0$, we can assume that the radius of the planetary magnetosphere corresponds to the radius of the planet, i.e., $R_{\rm m} = R_{\rm p}$ and apply, e.g., Eq.~(\ref{en_diss-nl}) simply substituting 
$B^{4/3} B_{p0}^{2/3} R_{\rm p}^{2}$ with $B^{2} R_{\rm p}^{2}$. 

When a planetary dipolar field is present, we find that the dissipated power scales as $R_{\rm p}^{2} B_{0}^{4/3} B_{p0}^{2/3}$, with $B_{0}$ the surface field of the star. This is a general result, independent of the specific mechanism responsible  for the magnetic energy dissipation, provided that the dissipated power is proportional to the available magnetic energy, that scales as $B_{0}^{2}$, and the surface of the planetary magnetosphere, that  scales as $B_{0}^{-2/3} B_{p0}^{2/3} R_{\rm p}^{2}$. This scaling law {may be used}  to infer the relative planetary field strength, not yet directly observable, from the measurement of the power of the SPMI in a sample of stars, as suggested by  \citet{Scharf10}.

An important conclusion of our model is that, while the reconnection of the stellar and planetary fields releases an additional power in the corona that can lead to an excess X-ray emission, this is not the case for the energy released  by the helicity variation that produces a modulation of the dissipation of the energy already available for the coronal heating or flaring. Since the power released by the reconnection is generally of the order of $10^{17}-10^{18}$~W, this can explain the recent results of \citet{Cantomartinsetal11} or those of \citet{Poppenhaegeretal10} that suggest at the most a marginal correlation of the stellar X-ray emission with the mass and the inverse orbital semimajor axis of the planet. On the other hand, a modulation of the coronal flaring activity with the orbit of the planet, as suggested by \citet{Pillitterietal11}, is predicted by our model.  Concerning the correlation found by \citet{Hartman10}, the range spanned by the  stellar chromospheric emission versus the planetary surface gravity  covers approximately one order of magnitude. This appears to be too large to be accounted for by the energy released by the interactions considered in our model and requires a different explanation.  

The time modulation of the chromospheric or X-ray emission predicted by our model has a single frequency, equal to the orbital frequency of the planet, only if the stellar magnetic field is predominantly axisymmetric, as expected when the star is close to the minimum of its activity cycle by analogy with the Sun. When the stellar field has a non-axisymmetric component comparable or larger than the axisymmetric one, the modulation becomes multiperiodic with frequencies coming from the combination of the orbital and stellar rotation frequencies and their harmonics, which makes the modulation virtually indistinguishable from the intrinsic activity fluctuations of the star. This may explain why in some seasons the chromospheric signature of the SPMI  has not been observed in stars that had previously shown some evidence of modulation with the orbital period. 

Models of the SPMI based on stellar linear force-free fields  allow us to treat both the case of axisymmetric and non-axisymmetric fields with an analytical description. However,  the power released, even assuming the extreme  values allowed for the free parameters, is insufficient by at least one order of magnitude. This is related to the existence of an upper bound  for the force-free parameter $\alpha$ because the field must extend at least up to the radius of the planetary orbit. On the other hand, such configurations can be useful to describe the final state of a confined magnetic structure, when its excess magnetic energy has been dissipated and the field has reached the minimum allowed energy compatible with the conservation of the relative magnetic helicity.   Such closed field configurations have been considered by, e.g., \citet{Lanza10} when discussing the evolution of the rotation of stars with hot Jupiters. 

 Non-linear force-free fields  in general do not have limitations on the value of the force-free parameter $\alpha$. However, the classes of fields that we have considered provide dissipated powers that are still insufficient by at least one order of magnitude to account for the observations, although they can account for the phase lag between the planet and the hot spot, with the exception of $\upsilon$~And. 
Therefore, our results cast doubts on a straightforward association of the observed chromospheric hot spots with the orbiting planets, requiring an alternative mechanism to explain the phenomenon.  Indeed, the numerical simulations of \citet{Cohenetal11b}, including also  the kinetic energy of the plasma flows in the
system, are more promising and could be more appropriate than the present idealized models to explain the observations. {In those numerical models, there are magnetic loops interconnecting the star with the planet. Therefore, the maximum power made available by the relative motion of the planet, as derived from the flux  of the Poynting vector $\mu^{-1} {\vec E}  \times {\vec B}$ across the base $\pi R_{\rm p}^{2}$ of an interconnecting flux tube, where $\vec E = - {\vec v} \times {\vec B}$ is the electric field and $\vec v$ the relative velocity, is: $P \simeq \pi \mu^{-1} R_{\rm p}^{2} B_{\rm p0}^{2} v$. For a relative velocity $ v = 10^{4}-10^{5}$~m~s$^{-1}$, a planetary field $B_{\rm p0} = 10$~G, and a  radius of the planet $R_{\rm p}= 7 \times 10^{7}$~m, we obtain a maximum available power of $\sim 10^{20}-10^{21}$~W that could be enough to account for the hot spots observed by \citet{Shkolniketal05}. We shall explore this possibility in more detail in a forthcoming work.  

}

\acknowledgements
The author is grateful to an anonymous Referee for a careful reading of the manuscript and valuable comments, and to Dr. E.~Shkolnik for several interesting discussions on star-planet interactions. Discussions with
Prof.~M.~Deleuil, Dr. C.~Moutou, Dr.~A.~S.~Bonomo, Dr.~I.~Pagano, Dr.~A.~Maggio, and Dr.~R.~Fares are also gratefully acknowledged. Research on active stars and exoplanets at INAF-Catania Astrophysical Observatory and the Dept. of Physics and Astronomy of the University of Catania are funded by MIUR ({\it Ministero dell'Istruzione, Universit\`a e Ricerca}).

\appendix
\section{The vector potential of a potential magnetic field}
\label{App1}
The application of Eq.~(\ref{dissipation_rate}) requires the determination of the vector potential ${\vec A}_{\rm p}$ for a given distribution of the normal component of the magnetic field $B_{\rm n} $ on the surface of the star. We consider a spherical polar coordinate system $(r, \theta, \phi )$ with the origin at the barycentre of the star and the polar axis along the rotation axis of the star. The surface of the star is a sphere of radius $R$. 

The normal component of the magnetic field at the surface is $B_{r} (R, \theta, \phi)$ and is equal to the radial component of  the potential magnetic field ${\vec B}_{\rm p} = \nabla \psi $, i.e., $(\partial \psi / \partial r)_{R} = B_{r} (R, \theta,\phi)$. The general expression of the scalar potential of the field ${\vec B}_{\rm p}$ exterior to the star is:
\begin{equation}
\psi (r, \theta, \phi ) = \sum_{n=1}^{\infty} \sum_{m=0}^{n} [a_{n m} Y_{n m}^{\rm e}(\theta, \phi) + b_{n m} Y_{n m}^{0} (\theta, \phi) ] \left( \frac{r}{R} \right)^{-(n+1)},
\label{psi-pot}
\end{equation}
where $a_{n m}$ and $b_{n m}$ are the numerical coefficients of the expansion of the scalar potential in terms of the 
spherical harmonic functions $Y_{nm}^{\rm e} (\theta, \phi) = P_{n}^{m} (\cos \theta) \cos (m \phi)$ and $Y_{nm}^{\rm 0} (\theta, \phi) = P_{n}^{m} (\cos \theta) \sin (m \phi)$, and $P_{n}^{m}$ are  Legendre associated polynomials of degree $n$ and azimuthal order $m$ with $0 \leq m \leq n$. The term corresponding to $n=0$ is absent in the development because the magnetic field has no monopoles. If the field is interior to the sphere of radius $R$, Eq.~(\ref{psi-pot}) is still valid provided that the radial dependence of the component  of the order $n$ is changed as $(r/R)^{n}$. 

Thanks to the orthogonality properties of the spherical harmonics, we can immediately derive the coefficients $a_{mn}$ and $b_{nm}$ from a surface integration of the radial field component, i.e.:
\begin{eqnarray}
a_{n m}  & = & \frac{1}{N_{mn}} \int_{0}^{2\pi} \int_{0}^{\pi} R B_{r} (R, \theta, \phi) Y_{nm}^{\rm e}(\theta, \phi) \sin \theta  \ d\theta  \ d \phi, \nonumber \\
b_{n m}  & = & \frac{1}{N_{mn}} \int_{0}^{2\pi} \int_{0}^{\pi} R B_{r} (R, \theta, \phi) Y_{nm}^{0}(\theta, \phi) \sin \theta  \ d\theta  \ d \phi, 
\label{pot-coeff}
\end{eqnarray}
where the normalization factor 
\begin{equation}
N_{n m} = - \ 4\pi \ \frac{(n+1)}{(2n+1)} \frac{(n+m)!}{(n-m)!}.
\end{equation}
 The connection between the scalar potential $\psi (r, \theta, \phi)$ and the vector potential ${\vec A}_{\rm p}$ can be derived as follows. In the representation of \citet{Chandrasekhar61}, a potential field has only a poloidal component, i.e., it can be written as: ${\vec B}_{\rm p}= \nabla \times
\left\{ \nabla \times [\Phi (r, \theta, \phi) \hat{\vec r}] \right\}$, where $\hat{\vec r}$ is the unit vector in the radial direction and $\Phi (r, \theta, \phi)$ a scalar function that will be specified below. Therefore,
the corresponding vector potential satisfying $\nabla \cdot {\vec A}_{\rm p}=0$ is: 
\begin{eqnarray}
{\vec A}_{\rm p} & = & \nabla \times [\Phi (r, \theta, \phi) \hat{\vec r}]  = \nonumber \\
 & = & \frac{1}{r \sin \theta} \left( \frac{\partial \Phi}{\partial \phi} \right) \hat{\vec \theta} - \frac{1}{r} \left( \frac{\partial \Phi}{\partial \theta} \right) \hat{\vec \phi}, 
 \label{ap-comp}
\end{eqnarray}
where $\hat{\vec \theta}$ and $\hat{\vec \phi}$ are the unit vectors in the meridional and the azimuthal directions, respectively. 
To derive the function $\Phi$, we compare the components of the field ${\vec B}_{\rm p}$ as given by the formulae of \citet{Chandrasekhar61} with the components of $\nabla \psi$. Chandrasekhar's formulae are:
\begin{equation}
B_{\rm p   r} = \frac{1}{r^{2}} L^{2} \Phi, \;  B_{\rm p  \theta} = \frac{1}{r} \frac{\partial }{\partial \theta} \left( \frac{\partial \Phi}{\partial r} \right), \;  B_{\rm p   \phi} = 
\frac{1}{ r \sin \theta} \frac{\partial}{\partial \phi} \left( \frac{\partial \Phi}{\partial r} \right),
\label{bp-poloidal}
\end{equation}
where 
\begin{equation}
L^{2} = -\frac{1}{\sin \theta} \frac{\partial}{\partial \theta} \left(  \sin \theta \frac{\partial}{\partial \theta} \right) - \frac{1}{\sin^{2} \theta} \frac{\partial^{2}}{\partial \phi^{2}}
\end{equation}
is the angular part of the Laplacian operator, i.e., $\nabla^{2} = \frac{1}{r^{2}} \frac{\partial }{\partial r} \left( r^{2} \frac{\partial}{\partial r} \right) - \frac{L^{2}}{r^{2}}$. Since the field ${\vec B}_{\rm p}$ is potential, $\nabla^{2} \psi = 0$. From this equation and the first of Eq.~(\ref{bp-poloidal}), it is easy to see that $\psi = \frac{\partial \Phi}{\partial r}$. The same expression immediately verifies also the second and the third relationships in~(\ref{bp-poloidal}); hence we find:
\begin{eqnarray}
\lefteqn{ \Phi (r, \theta, \phi)  =  \int \psi(r^{\prime}, \theta, \phi) \ dr^{\prime} = \nonumber } \\
  & = & -\sum_{n=1}^{\infty}   \frac{1}{n}  \sum_{m=0}^{n} R [a_{n m} Y_{n m}^{\rm e}(\theta, \phi) + b_{n m} Y_{n m}^{0} (\theta, \phi) ] \left( \frac{r}{R} \right)^{-n},
 \label{phi-expr}
\end{eqnarray}
where $a_{n m}$ and $b_{n m}$ are determined by the normal component of the magnetic field at the surface of the star according to Eqs.~(\ref{pot-coeff}). By substituing Eq.~(\ref{phi-expr}) into Eq.~(\ref{ap-comp}), we derive the components of the vector potential  ${\vec A}_{\rm p}$. Note that when the field is axisymmetric ($\partial / \partial \phi = 0$), ${\vec A}_{\rm p}$ has only the azimuthal component, i. e., it is everywhere parallel to the orbital velocity of the planet assumed to move on an equatorial and circular orbit. 

\section{Perturbation of the energy of the coronal field}
\label{field-energy}

The presence of the planetary magnetosphere produces a perturbation of the energy of the coronal field. If the planet were absent, the total energy of the field would be:
\begin{equation}
E_{0} = \frac{1}{2 \mu} \int_{V} B^{2} \ dV,
\end{equation}
where ${\vec B}$ is the coronal field and $V$ the volume exterior to the star. When the planet is present, the coronal field is ${\vec B}_{\rm m} = {\vec B} + {\vec B}^{\prime}$, where ${\vec B}^{\prime} = \nabla \zeta$ and the potential $\zeta$ is given by Eq.~(\ref{zeta-express}). Therefore, the energy of the coronal field becomes:
\begin{equation}
E_{1} = \frac{1}{2\mu} \int_{V^{\prime}} B_{\rm m}^{2} \ dV,
\end{equation}
where $V^{\prime}$ is the volume of the corona exterior to the star and the planetary magnetosphere. Since $B_{\rm m}^{2} = B^{2} + 2 {\vec B} \cdot {\vec B}^{\prime} + B^{\prime 2} = B^{2} +  ({\vec B} + {\vec B}^{\prime}) \cdot {\vec B}^{\prime} + {\vec B} \cdot {\vec B}^{\prime} = B^{2} + ({\vec B} + {\vec B}^{\prime}) \cdot \nabla \zeta + {\vec B} \cdot \nabla \zeta = B^{2} + {\vec B}_{\rm m} \cdot \nabla \zeta + {\vec B} \cdot \nabla \zeta = B^{2} + \nabla \cdot (\zeta { \vec B}_{\rm m}) + \nabla \cdot (\zeta {\vec B})$, we can apply Gauss' theorem to find: 
\begin{equation}
E_{1} = \frac{1}{2\mu} \left( \int_{V^{\prime}} B^{2} \ dV + \oint_{S_{\rm a} \cup S_{\rm m}} \zeta ({\vec B}_{\rm m} \cdot \hat{\vec n}) \ dS + \oint_{S_{\rm a} \cup S_{\rm m}} \zeta ({\vec B} \cdot \hat{\vec n}) \ dS \right),
\end{equation}
where $S_{\rm a}$ is the surface of the star and $S_{\rm m}$ the surface of the magnetosphere. The potential $\zeta = (1/2) ({\vec B} \cdot \hat{\vec r}) (R_{\rm m}^{3}/r^{2})$ is negligible on the surface of the star, while on the boundary of the magnetosphere ${\vec B}_{\rm m} \cdot \hat{\vec n} = 0$. Therefore, the middle integral vanishes and in the third integral only the integration over $S_{\rm m} $ remains, yielding:
\begin{equation}
E_{1} = \frac{1}{2\mu}  \int_{V^{\prime}} B^{2} \ dV  - \frac{1}{4 \mu} R_{\rm m} \oint_{S_{\rm m}} ({\vec B} \cdot \hat{\vec r})^{2} \ dS, 
\end{equation}
where the minus sign in front of the second integral comes from the orientation of the normal to $S_{\rm m}$, i.e., $\hat{\vec n} = - \hat{\vec r}$.
Adopting a reference frame with the $\hat{z}$ axis along the relative velocity $\vec v$ between the planet and the coronal field, and the $xz$ plane containing the vector $\vec B$, as specified in the text, we can perform the integration and find:
\begin{equation}
E_{1} = \frac{1}{2\mu}  \int_{V^{\prime}} B^{2} \ dV  - \frac{1}{4 \mu} V_{\rm m} B^{2}, 
\end{equation}
where $V_{\rm m}=(4\pi/3) R_{\rm m}^{3}$ is the volume of the magnetosphere. Since the radius of the magnetosphere is given by Eq.~(\ref{rm}), we recast this equation as:
\begin{equation}
E_{1} = \frac{1}{2\mu} \int_{V^{\prime}} B^{2} \ dV - \frac{\pi}{3 \mu} R_{\rm p}^{3} B_{\rm p0} B,
\end{equation}
where $R_{\rm p}$ is the radius of the planet and $B_{\rm p0}$ the planetary field at the pole. Finally, the available energy difference between the coronal field configurations with and without the planet is:
\begin{equation}
\Delta E \equiv E_{1} - E_{0} = - \frac{3}{4 \mu} V_{\rm m} B^{2} = - \frac{\pi}{\mu} R_{\rm p}^{3} B_{\rm p0} B,
\end{equation}
where we made use of $V = V^{\prime} \cup V_{\rm m}$, with $V_{\rm m}$  the volume of the magnetosphere, and the energy of the coronal field in that volume when the planet is absent is $\int_{V_{\rm m}} (B^{2}/2 \mu) \ dV = V_{\rm m} B^{2}/2\mu$, provided that $\vec B$ can be assumed uniform over the volume $V_{\rm m}$.

\section{Linear force-free field and  vector potential of the corresponding potential field}
\label{field-express}
 
The components of the linear force-free field of order $n=1$ as introduced by \citet{ChandrasekharKendall57} can be written as (see Appendix~\ref{App1} for the adopted reference frame): 
 \begin{eqnarray}
B_{r} & = &  \left( \frac{q_{0}}{q} \right)^{2}  g(q) \left[ B_{0 }\cos \theta + B_{1} \sin \theta \cos (\phi - \xi) \right], \nonumber \\
B_{\theta} & = & -\frac{1}{2} \frac{q_{0}^{2}}{q} \left\{ g^{\prime}(q) \left[  B_{0} \sin \theta - B_{1} \cos \theta \cos (\phi - \xi) \right]   + \right. \nonumber \\
 &  + & \left. g(q)  B_{1} \sin (\phi - \xi) \right\}  \nonumber \\
B_{\phi} & = & \frac{1}{2} \frac{q_{0}^{2}}{q}\left[ g(q) B_{0} \sin \theta +
 B_{1} g^{\prime} (q)  \sin (\phi - \xi) + \right. \nonumber \\
 & -  & \left. B_{1} g(q) \cos \theta \cos (\phi - \xi) \right], 
 \end{eqnarray} 
 where $B_{0}$ is the intensity of the axisymmetric mode of the field, i.e., that with $m=0$, at the poles and $B_{1}$ is the intensity of the radial component of the non-axisymmetric mode with $m=1$  at the longitude $\xi$. Note that the axisymmetric field has a zero radial component on the equatorial plane because  the field is similar to that of a dipole; the dimensionless radial distances $q \equiv |\alpha | r$ and $q_{0} \equiv | \alpha | R$, while the function $g$ is defined as:
 \begin{equation}
 g(q) \equiv \frac{q Z_{1}(q)}{q_{0} Z_{1}(q_{0})} = \frac{[b_{0} J_{-3/2}(q) + c_{0} J_{3/2}(q)]\sqrt{q}}{ [b_{0} J_{-3/2}(q_{0}) + c_{0} J_{3/2}(q_{0})]\sqrt{q_{0}} }, 
 \label{gq}
 \end{equation}
 where $Z_{1}$ is defined by Eq.~(10) of \citet{ChandrasekharKendall57} for $n=1$, $J_{-3/2}$ and $J_{3/2}$ are  Bessel functions of the first kind of order $-3/2$ and $3/2$, respectively, and $b_{0}$ and $c_{0}$ are coefficients determined by the field at the base of the corona \citep[see ][ for more details]{Lanza09}; the function $g^{\prime} \equiv dg/dq$ is the first derivative of $g$. When comparing the present expression of the magnetic field with that of \citet{ChandrasekharKendall57}, note that in the second line of their Eq.~(13) there is a typo, thus the correct argument of their  radial derivative is $[r Z_{n} (\alpha r)]$. 
 
 The components of the vector potential of the potential magnetic field having the same radial component of the force-free field at the surface, computed with the method introduced in Appendix~\ref{App1}, are:
 \begin{eqnarray}
 A_{\rm p r}  & = & 0, \nonumber \\
 A_{\rm p \theta } & = & -\frac{1}{2} R B_{1} \sin (\phi -\xi) \left(   \frac{q_{0}}{q} \right)^{2}, \nonumber \\
 A_{\rm p \phi} & = & \frac{1}{2} R \left[ B_{0} \sin \theta - B_{1} \cos \theta \cos (\phi - \xi) \right]  \left(   \frac{q_{0}}{q} \right)^{2}.
 \end{eqnarray}

\end{document}